\documentclass[11pt,a4paper]{article}
\usepackage[left=2.35cm,top=3cm,bottom=3cm,right=2.3cm]{geometry}
\usepackage{authblk}
\usepackage{amsmath}
\usepackage{amssymb}
\usepackage{tikz}
\usetikzlibrary{patterns}
\usetikzlibrary{patterns.meta}
\usepackage{float}
\usepackage{comment}
\usepackage{cite}
\usepackage{booktabs}
\usepackage{subcaption}
\usepackage[nolist,nohyperlinks]{acronym}
\usepackage[colorlinks=true
,urlcolor=blue
,anchorcolor=blue
,citecolor=blue
,filecolor=blue
,linkcolor=blue
,menucolor=blue
,linktocpage=true
,pdfproducer=medialab
]{hyperref}
\usepackage{orcidlink}
\usepackage{slashed}

\def\D{{\rm d}}

\def\sth{{4m_\pi^2}}
\def\dispint{\int_\sth^\infty \D \dpone}
\def\dispinttwo{\int_\sth^\infty \D \dptwo}

\def\Fpi{F_\pi^V}

\def\io{i\delta}
\def\thavg{\theta_{\rm avg}}
\def\eepp{$ ee\to \pi \pi$}
\def\eeppg{$ ee\to \pi \pi \gamma$}

\def\dpone{{s_1}}
\def\dptwo{{s_2}}
\def\dponesq{{s^2_1}}

\def\thrs{s_{\rm ths}}

\def\ieik{\hat{\mathcal{E}}}

\def\mcmule{{{\sc McMule}}}

\catcode`,\active

\newcommand{\problem}[2][blue]{%
  \fcolorbox{#1}{white}{$#2$}%
}
\newcommand{\beyond}[2][purple]{%
  \fcolorbox{#1}{white}{$#2$}%
}
\catcode`\,12

\definecolor{bluegray}{RGB}{140, 170, 250}
\definecolor{caribbeangreen}{rgb}{0.0, 0.8, 0.6}

\renewcommand{\Re}{\mathrm{Re}}
\renewcommand{\Im}{\mathrm{Im}}

\newcommand{\OpenLoops}{{\rmfamily\scshape OpenLoops}}
\newcommand{\vp}{\mathsf{p}}  

\usetikzlibrary{decorations.markings,decorations.pathmorphing,positioning,decorations.pathreplacing,shapes,calc}
\tikzset{
    photon/.style={decorate, decoration={snake,amplitude=1pt,segment length=6pt}},
    zigzag/.style={decorate, decoration=zigzag},
    tighterphoton/.style={
    decorate,
    decoration={
      snake,
      amplitude=1pt,
      segment length=3pt
    }}
}

\hypersetup{pdftitle={Radiative pion-pair production in disperon QED},
pdfauthor={
Yizhou Fang,
Sophie Kollatzsch,
Adrian Signer,
Max Zoller}}

\title{Radiative pion-pair production in disperon QED}
\date{}
\author[]{Yizhou~Fang\,\orcidlink{0009-0003-3568-9661}\,}
\author[]{Sophie~Kollatzsch\,\orcidlink{0000-0002-8560-1619}\,}
\author[]{Adrian~Signer\,\orcidlink{0000-0001-8488-7400}\,}
\author[]{Max~Zoller\,\orcidlink{0000-0001-9233-7951}\,}
\affil[]{PSI Center for Neutron and Muon Sciences, 5232 Villigen PSI, Switzerland}
\affil[]{Physik-Institut, Universit\"at Z\"urich, 8057 Z\"urich, Switzerland}

\allowdisplaybreaks

\begin{document}

\def\centerarc[#1](#2,#3)(#4:#5:#6)
    { \draw[#1] ({#2+#6*cos(#4)},{#3+#6*sin(#4)}) arc (#4:#5:#6); }

\newcommand\boxdiagphoton[4]{
        \begin{tikzpicture}[scale=#3,baseline=-0.5ex]
            \draw (135:0.7) -- (135:0.5) -- (-135:0.5) -- (-135:0.7);
            \draw[#4] (45:0.7) -- (45:0.5) -- (-45:0.5) -- (-45:0.7);
            \draw[#1] (135:0.5) -- (45:0.5);
            \draw[#2] (-135:0.5) -- (-45:0.5);
            \draw[photon] (-135:0.6) -- ++(-40:0.5);
        \end{tikzpicture}
}
\newcommand\boxdiagphotontwo[4]{
        \begin{tikzpicture}[scale=#3,baseline=-0.5ex]
            \draw (135:0.7) -- (135:0.5) -- (-135:0.5) -- (-135:0.7);
            \draw[#4] (45:0.7) -- (45:0.5) -- (-45:0.5) -- (-45:0.7);
            \draw[#1] (135:0.5) -- (45:0.5);
            \draw[#2] (-135:0.5) -- (-45:0.5);
            \draw[photon] (135:0.6) -- ++(40:0.5);
        \end{tikzpicture}
}
\newcommand\pentagon[4]{
        \begin{tikzpicture}[scale=#3,baseline=-0.5ex]
            \draw (135:0.7) -- (135:0.5) -- (-135:0.5) -- (-135:0.7);
            \draw[#4] (45:0.7) -- (45:0.5) -- (-45:0.5) -- (-45:0.7);
            \draw[#1] (135:0.5) -- (45:0.5);
            \draw[#2] (-135:0.5) -- (-45:0.5);
            \draw[photon] (-0.35, 0) -- ++(0:-0.4);
        \end{tikzpicture}
}

\definecolor{bluegray}{RGB}{140, 170, 250}

\makeatletter
    \newacro{DET}{disperon effective theory}
    \newacro{EFT}{effective field theory}
    \newacro{sQED}{scalar QED}
    \newacro{FsQED}{form-factor scalar QED}
    \newacro{FxsQED}[F$\times$sQED]{form-factor times scalar QED}
    \newacro{GVMD}{generalised vector-meson dominance}
    \newacro{HVP}{hadronic vacuum polarisation}
    \newacro{MC}{Monte Carlo}
    \newacro{MoR}{method of regions}
    \newacro{LO}{leading order}
    \newacro{NLO}{next-to-leading order\def\AC@acronyms@NNLO{@<>@<>@}\def\AC@acronyms@LO{@<>@<>@}\def\AC@acronyms@NNNLO{@<>@<>@}}
    \newacro{NNLO}{next-to-next-to-leading order\def\AC@acronyms@NNNLO{@<>@<>@}\def\AC@acronyms@NLO{@<>@<>@}\def\AC@acronyms@LO{@<>@<>@}}
    \newacro{NNNLO}[N$^3$LO]{next-to-next-to-next-to-leading order\def\AC@acronyms@NLO{@<>@<>@}\def\AC@acronyms@LO{@<>@<>@}\def\AC@acronyms@NNLO{@<>@<>@}}
    \newacro{VFF}{vector form factor}
    \newacro{VP}{vacuum polarisation}
    \newacro{dp}[{\tt dp}]{double precision}
    \newacro{qp}[{\tt qp}]{quadruple precision}
    \newacro{IR}{infrared}
    \newacro{UV}{ultraviolet}
    \newacro{ISC}{initial-state corrections}
    \newacro{FSC}{final-state corrections}
    \newacro{CT}{counterterm}
    \newacro{TPE}{two-photon exchange}
    \newacro{FKSl}[FKS$^\ell$]{}
    \newacro{NTS}{next-to-soft}
    \newacro{PS}{parton shower}
    \acrodefplural{CT}{counterterms}
\makeatother

\begin{titlepage}
\clearpage\maketitle 
\thispagestyle{empty}

\begin{abstract}\noindent
Based on the previously developed disperon QED method
for incorporating data input into loop processes in Monte Carlo
codes, we calculate \eeppg{} at NLO and include the pion vector form factor within the form-factor scalar QED approximation wherever it is physically justified. In particular, this includes insertions of the two-pion two-photon Compton tensor into loop diagrams. 
We also consider a subset of contributions beyond, involving the pion Compton tensor with three photons. For these terms the form-factor scalar QED approach is less well justified, but
we gain insight into the associated theory uncertainty for selected experimental scenarios.
All calculations are implemented and made available within the Monte Carlo framework \mcmule{}.
\end{abstract}
\end{titlepage}
\section{Introduction}
\label{sec:intro}
The anomalous magnetic moment of the muon $a_\mu$~\cite{Aliberti:2025beg} lies at the centre of several puzzles in particle physics.
Its \ac{HVP} contributions can be extracted from low-energy $e^+ \,e^-\to\,\text{hadrons}$ scattering experiments with pion-pair production providing the dominant contribution.
While energy scan experiments access the \ac{HVP} through measuring \eepp{} at various centre-of-mass energies $\sqrt{s}$, 
most experiments rely on radiative-return measurements of \eeppg{} at fixed $\sqrt{s}$.
However, significant tensions exist between different experimental determinations.
Recent lattice calculations~\cite{Borsanyi:2020mff,RBC:2023pvn,Djukanovic:2024cmq} add another puzzle: their \ac{HVP} contribution to $a_\mu$ results in a  theoretical prediction that is in agreement with the experimental measurement of $a_\mu$~\cite{Muong-2:2025xyk}, but differs from most data-driven determinations.

Resolving these tensions requires a precise understanding of the measurements of 
hadronic cross sections and, consequently, on the theoretical description of the underlying scattering processes.
For $e^+ \,e^-\to\,\text{hadrons}$ measurements at fixed $\sqrt{s}$, these are mostly predictions for $ee\to \mu\mu\gamma$ and $ee\to\pi\pi\gamma$, implemented into a fully differential Monte Carlo tool~\cite{Aliberti:2024fpq}.
For $ee\to \mu\mu \gamma$, QED corrections are available at \ac{NLO} matched to a parton shower~\cite{Budassi:2026lmr}.
Work towards a \ac{NNLO} prediction is ongoing~\cite{Fadin:2023phc,Badger:2023xtl,PetitRosas:2025xhm,Pozzoli:2026eiu,Kollatzsch:2026TI,Fang:2026dbs}.

For \eeppg{} however, the situation is more involved. 
Due to the composite structure of the pion, the $\gamma^*\to\pi\pi$ interaction vertex includes a non-perturbative \ac{VFF}  $F^V_\pi$.
Additional hadronic matrix elements such as Compton tensors $(\gamma^*)^{2+k} \to \pi\pi$ have to be included as well in the amplitude. 
This leads to technical and conceptual challenges.

The technical challenge is to bring the complicated hadronic input into a form that is compatible with the loop integration in \eeppg{}.
For pion-photon interactions, it has been shown that using \ac{sQED} equipped with a pion \ac{VFF} at each vertex, is a reasonable first step to model the Compton tensors required for \eeppg{}~\cite{Colangelo:2015ama,Aliberti:2024fpq,Kaziukenas:2025gggpp}.
If the Compton tensor is expressed in this way, this is referred to as \ac{FxsQED}.
If the \ac{VFF} entering this prescription is parametrised in terms of standard loop propagators, this approach is referred to as \ac{GVMD}~\cite{Sakurai:1972wk,Ignatov:2022iou}.
This comes at the cost of giving up analyticity of the \ac{VFF}, and complications in restoring
unitarity.
Recently, predictions for \eeppg{} in \ac{GVMD} became available at fixed \ac{NLO}~\cite{PetitRosas:2026iuq} and including a parton shower~\cite{CarloniCalame:2026vfc} or all-order soft photon exponentiation~\cite{Delorme:2026qbm}.
Alternatively, if the \ac{VFF} in a general, for example numerical, representation is incorporated into the loop integral, the result is referred to as (the technical definition of) \ac{FsQED}.
This can be achieved by applying a dispersion relation that avoids the evaluation of $F^V_\pi$ for the loop momentum but moves it inside an additional dispersive integral~\cite{Colangelo:2014dfa,Colangelo:2015ama,Colangelo:2022lzg}.
The latter can be carried out numerically.
As already emphasised in~\cite{Fang:2025mhn}, \ac{FsQED} was originally defined in~\cite{Aliberti:2024fpq} as including the pole terms of the Compton tensors.
While this coincides with the technical definition in the case of the $\gamma^*\gamma^*\to\pi\pi$ Compton tensor, we generalise this prescription and refer to \ac{FsQED} whenever pion-photon interactions are described by \ac{FxsQED} with a suitable pion \ac{VFF}.

Predictions for \eeppg{}, where (the technical version of) \ac{FsQED} was used everywhere, are available at parton shower improved \ac{NLO}~\cite{CarloniCalame:2026gdm}, where no significant deviations from the \ac{GVMD} result~\cite{CarloniCalame:2026vfc} were reported.
Together, these approaches now supersede the previously used \ac{FxsQED}
approach for the \eeppg{} amplitude~\cite{Czyz:2003ue,Tracz:2018,Campanario:2019mjh}, where pion \ac{VFF} only appear externally outside the loop integral.

The conceptual challenges are related to the question which part of \ac{NLO} predictions of \eeppg{}  can be reliably approximated through \ac{FsQED}, i.e. by the insertion of \ac{FxsQED} blocks.
For this discussion it is very convenient to split contributions into \ac{ISC}, \ac{FSC} and mixed corrections. 
The \ac{ISC} only contain radiative corrections associated to the electron line in \eeppg{}.
Here, only the pion \ac{VFF}, multiplied externally, is required.
\ac{FSC} on the other hand describe corrections exclusively on the pion side.
Given their hadronic structure, they are extremely difficult~\cite{Aliberti:2024fpq} and \ac{FsQED} -- while technically feasible -- might not provide a reliable prediction.
Mixed corrections describe interactions and interference effects between initial and finite states.
They contain topologies, where Compton tensors $(\gamma^*)^{2+k} \to \pi\pi$ and radiative corrections thereof need to be included and a careful analysis is required to investigate for which parts \ac{FsQED} is a reliable approximation.
Aside from the separate challenges associated to the different categories, the classification also gives a first hint on their numerical impact, as \ac{ISC} may be dominant due to the logarithmic enhancement associated with $m_e \ll m_\pi$ and the form-factor suppression of final-state radiation since $F^V_\pi(q^2 \to \infty)\to 0$.

Upcoming analyses of \eeppg{}, such as the ongoing KLOE-NXT analysis~\cite{Punzi:2026TI}, require a precision of at least $0.1\%$ for differential cross sections for \eeppg{}~\cite{KLOE:private}.
Meeting this target demands a substantial effort on the theory side, 
both in understanding the underlying non-perturbative structure of the pion-photon interactions~\cite{Hoferichter:2013ama,Colangelo:2015ama,Hoferichter:2019nlq,Ludtke:2023hvz,Kaziukenas:2025gggpp,Lymperiadou:2026gggpp,Monnard:2021pvm,Aliberti:2024fpq,Colangelo:2022lzg} as well as the connection of such methods with Monte Carlo codes~\cite{Aliberti:2024fpq,Hoferichter:2025rescatteringpp,Budassi:2024whw,Fang:2025mhn,Budassi:2026lmr,CarloniCalame:2026vfc,CarloniCalame:2026gdm}.
This work is devoted to the latter task. In particular, we focus on the subset of mixed corrections at \ac{NLO} that 
can be reliably calculated using the \ac{FsQED} prescription. 
This will be combined with
the \ac{NLO} \ac{ISC} to \eeppg{} which have already been presented in~\cite{Aliberti:2024fpq} by several codes, including \mcmule{}.
As we will argue, however, \ac{FsQED} does not provide a physically justifiable description for \ac{FSC}.

The results presented in this work are part of the Monte Carlo framework \mcmule{}~\cite{Banerjee:2020rww, McMule:manual} whose code (publicly available version \texttt{v.0.8.0}) and results are publicly available at
\begin{quote}
    \url{https://mule-tools.gitlab.io}
\end{quote}
They rely on disperon QED~\cite{Fang:2025mhn}, a method that has been developed to deal with complicated data input in general loop processes in Monte Carlo tools, as it is the case for \ac{FsQED}.

This work is structured as follows.
First in Sections~\ref{sec:compton} and~\ref{sec:disperonQED}, we focus on the challenges of \ac{FsQED}, both from a conceptual (Section~\ref{sec:compton}) and technical (Section~\ref{sec:disperonQED}) point of view.
In particular, in Section~\ref{sec:compton}, we summarise the fundamental foundations the \ac{FsQED} description is based upon and comment on its limitations. 
This leads us to define a subset of \ac{NLO} corrections to \eeppg{} that can be reliably calculated using \ac{FsQED}.
Based on this, we are ready to show phenomenological results in Section~\ref{sec:results} before concluding and commenting on future work in Section~\ref{sec:conclusions}. 
\section{Conceptual challenges of FsQED}
\label{sec:compton}
Amplitudes involving hadronic final states depend on complicated objects that describe the interaction of hadrons with photons.
For \ac{ISC} in \eeppg{}, the pion \ac{VFF} $F^V_\pi(q^2)$ defined as the interaction of a photon carrying momentum $q^2$ with two on-shell charged pions is sufficient.
For mixed and \ac{FSC}, the amplitudes $\gamma^* \gamma^* \to \pi\pi$ and $\gamma^* \gamma^* \gamma \to \pi\pi$ are required.
Their description can be constructed from their discontinuities, intermediary on-shell states, by using dispersion relations~\cite{Colangelo:2015ama,Hoferichter:2019nlq,Colangelo:2017fiz}.
The first intermediary state is the pion itself.
For $\gamma^* \gamma^* \to \pi \pi$, this pion-pole contribution is given by a \ac{FxsQED} description of the Compton tensor.
Once inserted into the amplitude for $ee\to\pi\pi(\gamma)$, this results in the \ac{FsQED} description~\cite{Colangelo:2014dfa,Colangelo:2015ama,Colangelo:2022lzg}.
Non-pole terms and rescattering corrections between the pions are expected to be small~\cite{Aoyama:2020ynm}.
For $\gamma^* \gamma^* \gamma \to \pi\pi$ the situation is more complicated.
Here, rescattering corrections can be enhanced, and even the notion of a pion pole becomes ambiguous.
Nevertheless, using \ac{FxsQED} for $\gamma^* \gamma^* \gamma \to \pi\pi$ provides a reasonable starting point for the pion-pole contribution.
The different hadronic tensors that enter \eeppg{} together with the status of their dispersive description are summarised in Table~\ref{tab:ComptonTensors}.
For more details we refer to~\cite{Fang:2025mhn,Aliberti:2024fpq} and references therein. 

\begin{table}[]
\centering
\begin{tabular}{p{1.5cm} p{3.5cm} p{4.5cm} p{3cm}}
\toprule
hadronic tensor        &         $\begin{gathered}
    \begin{tikzpicture}
            \draw[dashed] (0, 0.5) -- (  45:1) node [right] {$\pi^+$};
            \draw[dashed] (0,-0.5) -- ( -45:1) node [right] {$\pi^-$};
            \draw[photon] (0, 0.5) -- ( 135:1) node [left ] {$\gamma^*$};
            \draw[photon] (0,-0.5) -- (-135:1) node [left ] {$\gamma^*$};
            \fill[gray] (0,0.) ellipse (0.3 and 0.7);
        \end{tikzpicture}\end{gathered}$ & 
        $\begin{gathered}
    \begin{tikzpicture}
            \draw[dashed] (0, 0.5) -- (  45:1) node [right] {$\pi^+$};
            \draw[dashed] (0,-0.5) -- ( -45:1) node [right] {$\pi^-$};
            \draw[photon] (0, 0.5) -- ( 135:1) node [left ] {$\gamma^*$};
            \draw[photon] (0, 0) -- ( 180:1) node [left ] {$\gamma$};
            \draw[photon] (0,-0.5) -- (-135:1) node [left ] {$\gamma^*$};
            \fill[gray] (0,0.) ellipse (0.3 and 0.7);
        \end{tikzpicture}\end{gathered}$
        &  
        $\begin{gathered}
        \begin{tikzpicture}
            \draw[photon] (0.5,0) -- (-0.5,0) node [left] {$\gamma^*$};
            \draw[tighterphoton] (0.9,0.3) arc[start angle=90,end angle=-90,radius=0.3];
            \draw[dashed] (0.5,0) -- (1.3,-0.8) node [right] {$\pi^-$};
            \draw[dashed] (0.5,0) -- (1.3,0.8) node[right] {$\pi^+$};
            \fill[gray]
        (0.6,0) ellipse (0.35 and 0.7);
        \end{tikzpicture}
    \end{gathered}$
        \\
        & $\gamma^* \gamma^* \to \pi\pi$& $\gamma^* \gamma^* \gamma \to \pi\pi$ & radiative corrections to Compton tensor \\
\midrule
status       
& pole contribution described by \ac{FxsQED}, full expression known~\cite{Hoferichter:2013ama,Colangelo:2015ama,Hoferichter:2019nlq} but too complicated for use in $e^+ e^-$ predictions~\cite{Hoferichter:2025rescatteringpp}
& notion of pole terms ambiguous, \ac{FxsQED} reasonable starting point for pole contribution if on-shell photon is soft~\cite{Ludtke:2023hvz,Kaziukenas:2025gggpp}, beyond-pole contributions required but not yet available, e.g.~\cite{Lymperiadou:2026gggpp}  
& could be calculated using \ac{sQED}, e.g.~\cite{Monnard:2021pvm}, under investigation~\cite{Aliberti:2024fpq} 
but physics justification questionable\\
\bottomrule
\end{tabular}
\caption{Pion Compton tensors and the status of their dispersive description.}
\label{tab:ComptonTensors}
\end{table}

For the process \eepp{},  for the whole set of mixed corrections the technical version of \ac{FsQED} coincides with the original one, describing the pole terms of the Compton tensors.
In fact, only the insertion of the Compton tensor $\gamma^*\gamma^*\to \pi\pi$ is required~\cite{Colangelo:2022lzg}.
For \eeppg{}, the situation is more involved. 
In the remainder of this section we discuss the different contributions and how they can be estimated.

\subsection{Charge counting and
\texorpdfstring{$q_e^5\,q_\pi^3$}{qe^5 qpi^3}
contribution}
To disentangle the mixed corrections for \eeppg{} further into gauge-invariant and \ac{IR} finite subsets, we first introduce a charge counting by assigning (formally) different charges to electron ($q_e$) and pion ($q_\pi$).
\begin{figure}[]
  \centering
  \begin{subfigure}{0.95\textwidth}
    \centering
    \begin{minipage}{0.85\textwidth}
  \begin{align*}
    &\left(\,
    \begin{gathered}
        \begin{tikzpicture}
            \draw (-1.3,0.8) -- (-0.5,0) -- (-1.3,-0.8);
            \draw[photon] (-0.5,0) -- (0.5,0);
            \draw[dashed] (1.3,0.8) -- (0.5,0) -- (1.3,-0.8);
            \draw[photon] (-0.9,0.4) -- (0.2,0.4);
            \fill[gray] (0.5,0.) circle (0.2);
        \end{tikzpicture}
    \end{gathered}
    \,\right)^\dagger
    \cdot
    \left(\,
    \begin{gathered}
        \begin{tikzpicture}
            \draw (-1.3,0.8) -- (-0.5,0) -- (-1.3,-0.8);
            \draw[photon] (-0.5,0) -- (0.5,0);
            \draw[photon] (-1.1,0.6) -- (-1.1,-0.6);
            \draw[dashed] (1.3,0.8) -- (0.5,0) -- (1.3,-0.8);
            \draw[photon] (0.9,0.4) -- (-0.2,0.4);
            \fill[fill=gray]
        (0.7,0) ellipse (0.42 and 0.7);
        \end{tikzpicture}
    \end{gathered}
    +
    \begin{gathered}
        \begin{tikzpicture}
            \draw (135:1.5) -- (135:0.9) -- (-135:0.9) -- (-135:1.5);
            \draw[dashed] (45:1.5) -- (45:0.9) -- (-45:0.9) -- (-45:1.5);
            \draw[photon] (135:0.9) -- (45:0.9);
            \draw[photon] (-135:0.9) -- (-45:0.9);
            \draw[photon] (-0.65,0) -- (-1.2,0.);
            \fill[fill=gray]
        (0.62,0) ellipse (0.42 and 0.8);
        \end{tikzpicture}
    \end{gathered}\,\right)\notag\\
    +&
    \left(\,
    \begin{gathered}
        \begin{tikzpicture}
            \draw (-1.3,0.8) -- (-0.5,0) -- (-1.3,-0.8);
            \draw[photon] (-0.5,0) -- (0.5,0);
            \draw[dashed] (1.3,0.8) -- (0.5,0) -- (1.3,-0.8);
            \draw[photon] (0.9,0.4) -- (-0.2,0.4);
            \fill[fill=gray]
        (0.7,0) ellipse (0.42 and 0.7);
        \end{tikzpicture}
    \end{gathered}
    \,\right)^\dagger
    \cdot
    \left(\,
    \begin{gathered}
        \begin{tikzpicture}
            \draw (-1.3,0.8) -- (-0.5,0) -- (-1.3,-0.8);
            \draw[photon] (-0.5,0) -- (0.5,0);
            \draw[photon] (-1.1,0.6) -- (-1.1,-0.6);
            \draw[dashed] (1.3,0.8) -- (0.5,0) -- (1.3,-0.8);
            \draw[photon] (-0.9,0.4) -- (0.2,0.4);
            \fill[gray] (0.5,0.) circle (0.2);
        \end{tikzpicture}
    \end{gathered}
    \,\right)
    \oplus\,
    \left(\,\begin{gathered}
        \begin{tikzpicture}
            \draw (-1.3,0.8) -- (-0.5,0) -- (-1.3,-0.8);
            \draw[photon] (-0.5,0) -- (0.5,0);
            \draw[photon] (-1.1,0.6) -- (0.1, 0.6);
            \draw[dashed] (1.3,0.8) -- (0.5,0) -- (1.3,-0.8);
            \draw[photon] (-0.8,0.3) -- (0.2,0.3);
            \fill[gray] (0.5,0.) circle (0.2);
        \end{tikzpicture}
    \end{gathered}\,\right)^\dagger
    \cdot
    \begin{gathered}
        \begin{tikzpicture}
            \draw (-1.3,0.8) -- (-0.5,0) -- (-1.3,-0.8);
            \draw[photon] (-0.5,0) -- (0.5,0);
            \draw[photon] (0.8,0.3) -- (-0.2, 0.3);
            \draw[dashed] (1.3,0.8) -- (0.5,0) -- (1.3,-0.8);
            \draw[photon] (-1.1,0.6) -- (0.1,0.6);
            \fill[fill=gray]
        (0.7,0) ellipse (0.39 and 0.65);
        \end{tikzpicture}
    \end{gathered}
  \end{align*}
  \end{minipage}
    \caption{$q_e^5 \,q_\pi^3$}
    \label{fig:e5p3}
  \end{subfigure}
~
  \begin{subfigure}{0.95\textwidth}
    \centering
      \begin{minipage}{0.85\textwidth}
  \begin{align*}
&\beyond{\left(\,
    \begin{gathered}
        \begin{tikzpicture}
            \draw (-1.3,0.8) -- (-0.5,0) -- (-1.3,-0.8);
            \draw[photon] (-0.5,0) -- (0.5,0);
            \draw[dashed] (1.3,0.8) -- (0.5,0) -- (1.3,-0.8);
            \draw[photon] (-0.9,0.4) -- (0.2,0.4);
            \fill[gray] (0.5,0.) circle (0.2);
        \end{tikzpicture}
    \end{gathered}
    \,\right)^\dagger
    \cdot
    \left(\,
    \begin{gathered}
        \begin{tikzpicture}
            \draw (135:1.5) -- (135:0.9) -- (-135:0.9) -- (-135:1.5);
            \draw[dashed] (45:1.5) -- (45:0.9) -- (-45:0.9) -- (-45:1.5);
            \draw[photon] (135:0.9) -- (45:0.9);
            \draw[photon] (-135:0.9) -- (-45:0.9);
            \draw[photon] (0.65,0) -- (1.2,0.);
            \fill[fill=gray]
        (0.6,0) ellipse (0.39 and 0.8);
        \end{tikzpicture}
    \end{gathered}\right.} \left.
    +
    \problem{\begin{gathered}
        \begin{tikzpicture}
            \draw (-1.3,0.8) -- (-0.5,0) -- (-1.3,-0.8);
            \draw[photon] (-0.5,0) -- (0.5,0);
            \draw[dashed] (1.3,0.8) -- (0.5,0) -- (1.3,-0.8);
            \draw[photon] (-0.9,0.4) -- (0,0.4);
            \draw[photon] (1,0.5) arc[start angle=90,end angle=-90,radius=0.5];
            \fill[fill=gray]
        (0.8,0) ellipse (0.39 and 0.8);
        \end{tikzpicture}
    \end{gathered}}\,\right)\notag\\
    +&
    \problem{
    \left(\,
    \begin{gathered}
        \begin{tikzpicture}
            \draw (-1.3,0.8) -- (-0.5,0) -- (-1.3,-0.8);
            \draw[photon] (-0.5,0) -- (0.5,0);
            \draw[dashed] (1.3,0.8) -- (0.5,0) -- (1.3,-0.8);
            \draw[photon] (0.9,0.4) -- (-0.2,0.4);
            \fill[fill=gray]
        (0.7,0) ellipse (0.42 and 0.7);
        \end{tikzpicture}
    \end{gathered}
    \,\right)^\dagger
    \cdot
    \left(\,
    \begin{gathered}
        \begin{tikzpicture}
            \draw (-1.3,0.8) -- (-0.5,0) -- (-1.3,-0.8);
            \draw[photon] (-0.5,0) -- (0.5,0);
            \draw[photon] (-1.1,0.6) -- (-1.1,-0.6);
            \draw[dashed] (1.3,0.8) -- (0.5,0) -- (1.3,-0.8);
            \draw[photon] (0.9,0.4) -- (-0.2,0.4);
            \fill[fill=gray]
        (0.6,0) ellipse (0.39 and 0.8);
        \end{tikzpicture}
    \end{gathered}\right.} \left.
    +
    \begin{gathered}
        \begin{tikzpicture}
            \draw (135:1.5) -- (135:0.9) -- (-135:0.9) -- (-135:1.5);
            \draw[dashed] (45:1.5) -- (45:0.9) -- (-45:0.9) -- (-45:1.5);
            \draw[photon] (135:0.9) -- (45:0.9);
            \draw[photon] (-135:0.9) -- (-45:0.9);
            \draw[photon] (-0.65,0) -- (-1.2,0.);
            \fill[fill=gray]
        (0.6,0) ellipse (0.39 and 0.8);
        \end{tikzpicture}
    \end{gathered}
    \,\right)
    \\
    \oplus&
    \beyond{\left(\,\begin{gathered}
        \begin{tikzpicture}
            \draw (-1.3,0.8) -- (-0.5,0) -- (-1.3,-0.8);
            \draw[photon] (-0.5,0) -- (0.5,0);
            \draw[photon] (-1.1,0.6) -- (0.1, 0.6);
            \draw[dashed] (1.3,0.8) -- (0.5,0) -- (1.3,-0.8);
            \draw[photon] (-0.8,0.3) -- (0.2,0.3);
            \fill[gray] (0.5,0.) circle (0.2);
        \end{tikzpicture}
    \end{gathered}\,\right)^\dagger
    \cdot
    \begin{gathered}
        \begin{tikzpicture}
            \draw (-1.3,0.8) -- (-0.5,0) -- (-1.3,-0.8);
            \draw[photon] (-0.5,0) -- (0.5,0);
            \draw[photon] (0.8,-0.3) -- (-0.2, -0.3);
            \draw[dashed] (1.3,0.8) -- (0.5,0) -- (1.3,-0.8);
            \draw[photon] (0.8,0.3) -- (-0.2,0.3);
            \fill[fill=gray]
        (0.7,0) ellipse (0.42 and 0.7);
        \end{tikzpicture}
    \end{gathered}}
    \oplus
    \problem{\left|\,\,\begin{gathered}
        \begin{tikzpicture}
            \draw (-1.3,0.8) -- (-0.5,0) -- (-1.3,-0.8);
            \draw[photon] (-0.5,0) -- (0.5,0);
            \draw[photon] (0.8,0.3) -- (-0.2, 0.3);
            \draw[dashed] (1.3,0.8) -- (0.5,0) -- (1.3,-0.8);
            \draw[photon] (-1.1,0.6) -- (0.1,0.6);
            \fill[fill=gray]
        (0.7,0) ellipse (0.39 and 0.65);
        \end{tikzpicture}
    \end{gathered}\,\,\right|^2}
  \end{align*}
  \end{minipage}
    \caption{$q_e^4 \,q_\pi^4$}
    \label{fig:e4p4}
  \end{subfigure}
  \begin{subfigure}{0.85\textwidth}
    \centering
    \begin{align*}
    &\problem{\left(\,
   \begin{gathered}
        \begin{tikzpicture}
            \draw (-1.3,0.8) -- (-0.5,0) -- (-1.3,-0.8);
            \draw[photon] (-0.5,0) -- (0.5,0);
            \draw[dashed] (1.3,0.8) -- (0.5,0) -- (1.3,-0.8);
            \draw[photon] (0.9,0.4) -- (-0.2,0.4);
            \fill[fill=gray]
        (0.7,0) ellipse (0.42 and 0.7);
        \end{tikzpicture}
    \end{gathered}
    \,\right)^\dagger
    \cdot
\left(\,
    \begin{gathered}
        \begin{tikzpicture}
            \draw (135:1.5) -- (135:0.9) -- (-135:0.9) -- (-135:1.5);
            \draw[dashed] (45:1.5) -- (45:0.9) -- (-45:0.9) -- (-45:1.5);
            \draw[photon] (135:0.9) -- (45:0.9);
            \draw[photon] (-135:0.9) -- (-45:0.9);
            \draw[photon] (0.65,0) -- (1.2,0.);
            \fill[fill=gray]
        (0.6,0) ellipse (0.39 and 0.8);
        \end{tikzpicture}
    \end{gathered}\right.} \left.
    +
    \problem{\begin{gathered}
        \begin{tikzpicture}
            \draw (-1.3,0.8) -- (-0.5,0) -- (-1.3,-0.8);
            \draw[photon] (-0.5,0) -- (0.5,0);
            \draw[dashed] (1.3,0.8) -- (0.5,0) -- (1.3,-0.8);
            \draw[photon] (-0.9,0.4) -- (0,0.4);
            \draw[photon] (1,0.5) arc[start angle=90,end angle=-90,radius=0.5];
            \fill[fill=gray]
        (0.8,0) ellipse (0.39 and 0.8);
        \end{tikzpicture}
    \end{gathered}}\,\right)\\
    +&\left(\,
    \begin{gathered}
        \begin{tikzpicture}
            \draw (-1.3,0.8) -- (-0.5,0) -- (-1.3,-0.8);
            \draw[photon] (-0.5,0) -- (0.5,0);
            \draw[dashed] (1.3,0.8) -- (0.5,0) -- (1.3,-0.8);
            \draw[photon] (-0.9,0.4) -- (0.2,0.4);
            \fill[gray] (0.5,0.) circle (0.2);
        \end{tikzpicture}
    \end{gathered}
    \,\right)^\dagger
    \cdot
    \left(\,
    \problem{\begin{gathered}
        \begin{tikzpicture}
            \draw (-1.3,0.8) -- (-0.5,0) -- (-1.3,-0.8);
            \draw[photon] (-0.5,0) -- (0.5,0);
            \draw[dashed] (1.3,0.8) -- (0.5,0) -- (1.3,-0.8);
            \draw[photon] (0.9,0.4) -- (-0.2,0.4);
            \draw[photon] (1,0.5) arc[start angle=90,end angle=-90,radius=0.5];
            \fill[fill=gray]
        (0.8,0) ellipse (0.39 and 0.8);
        \end{tikzpicture}
    \end{gathered}}\,\right)\notag \oplus
    \problem{\left(\,\begin{gathered}
        \begin{tikzpicture}
            \draw (-1.3,0.8) -- (-0.5,0) -- (-1.3,-0.8);
            \draw[photon] (-0.5,0) -- (0.5,0);
            \draw[photon] (0.8,0.3) -- (-0.2, 0.3);
            \draw[dashed] (1.3,0.8) -- (0.5,0) -- (1.3,-0.8);
            \draw[photon] (-1.1,0.6) -- (0.1,0.6);
            \fill[fill=gray]
        (0.7,0) ellipse (0.39 and 0.65);
        \end{tikzpicture}
    \end{gathered}\,\right)^\dagger
    \cdot
\begin{gathered}
        \begin{tikzpicture}
            \draw (-1.3,0.8) -- (-0.5,0) -- (-1.3,-0.8);
            \draw[photon] (-0.5,0) -- (0.5,0);
            \draw[photon] (0.8,-0.3) -- (-0.2, -0.3);
            \draw[dashed] (1.3,0.8) -- (0.5,0) -- (1.3,-0.8);
            \draw[photon] (0.8,0.3) -- (-0.2,0.3);
            \fill[fill=gray]
        (0.7,0) ellipse (0.42 and 0.7);
        \end{tikzpicture}
    \end{gathered}}
    \notag
\end{align*}
    \caption{$q_e^3 \,q_\pi^5$}
    \label{fig:e3p5}
  \end{subfigure}
  \caption{Schematic overview of the individual subsets contributing to the mixed corrections at \ac{NLO} for \eeppg{}.
  Elliptic gray objects denote Compton tensors and radiative corrections thereof.
  Pion \acp{VFF} are denoted by gray circles.
  Only representative diagrams are shown.
  The problematic contributions that involve radiative corrections to Compton tensors are highlighted in blue.
  Contributions that can be tackled in \ac{FsQED} and go beyond $q_e^5 \,q_\pi^3$ are highlighted in purple. 
  The addition of real corrections is indicated with the symbol $\oplus$.
  }
  \label{fig:overviewmixed}
\end{figure}
The \ac{ISC} corrections scale on the squared-amplitude level as $q^6_e\,q^2_\pi$ and hence, the \ac{FSC} as $q^2_e\,q^6_\pi$.

The mixed corrections are divided into three subclasses, see Figure~\ref{fig:overviewmixed}, where we also show (using the symbol $\oplus$) the corresponding diagrams with additional real photon emission that are required for the \ac{IR} cancellation.
The charge counting that is assigned to each Compton tensor is based on its number of photon interactions.
A cancellation of \ac{IR} singularities requires all diagrams within a given subclass to be combined.
For the first subclass, $q_e^5 \,q_\pi^3$ shown in Figure~\ref{fig:e5p3}, this is straightforward since its calculation only requires the knowledge of $\gamma^*\gamma^*\to \pi\pi$ as it is the case for \eepp{}.

Already the next subclass, $q_e^4 \,q_\pi^4$ shown in Figure~\ref{fig:e4p4}, however, requires radiative corrections to the Compton tensor. 
For those effects, as discussed before and summarised in Table~\ref{tab:ComptonTensors}, \ac{FsQED} does not appear to be a suitable description. 
Hence, strictly speaking and following our previous discussion, this subclass should not be calculated in \ac{FsQED}.

It follows that \ac{ISC} and the $q_e^5\,q_\pi^3$ subset of the mixed corrections are the only gauge-invariant and IR finite subsets of the full set of \ac{NLO} corrections that can be calculated reliably in \ac{FsQED}.
In fact, for this subset, the technical definition of \ac{FsQED} coincides with the original one~\cite{Aliberti:2024fpq}.
A way of estimating the impact of contributions that go beyond that is described in the following.

\subsection{Contributions beyond 
\texorpdfstring{$q_e^5\,q_\pi^3$}{qe^5 qpi^3}
}
For the subset $q_e^4 \,q_\pi^4$, we have three contributions that challenge \ac{FsQED} according to Table~\ref{tab:ComptonTensors}.
They are highlighted in blue in Figure~\ref{fig:e4p4}.
They are either directly (blue highlighted diagram in the first line) or after evaluating their interference (blue highlighted diagrams in the second and third line) related to radiative corrections of the Compton tensor.

As mentioned before, it is strictly speaking not possible to separate the remaining contributions of $q_e^4 \,q_\pi^4$ from those mentioned above as they are connected through the \ac{IR} cancellation.
In order to still 
estimate the impact of $q_e^4 \,q_\pi^4$ on physical
observables, we introduce a hadronic counting parameter $\lambda_{\text{hadr}}$ together with a set of rules that makes a further split of $q_e^4 \,q_\pi^4$ possible.
\begin{itemize}
    \item[1)] We start with the diagrams in Figure~\ref{fig:e4p4}, where all interactions are expressed in terms of \acp{VFF} and Compton tensors. 
    \item[2)] Every Compton tensor scales as $\lambda^1_{\text{hadr}}$. 
    The pion \ac{VFF} scales as $\lambda^0_{\text{hadr}}$.
    \item[3)] We use two selection rules. 
    The selected diagrams are highlighted in purple in Figure~\ref{fig:e4p4}.
    \begin{itemize}
        \item[a)] We only take contributions that scale as at most $\lambda^1_{\text{hadr}}$ on their squared-amplitude level.
    Contributions with $\lambda^2_{\text{hadr}}$ and higher are dropped.
    \item[b)] Radiative corrections to Compton tensors (such as the blue highlighted diagram in the first line of Figure~\ref{fig:e4p4}) are dropped.
    \end{itemize}
    \item[4)] Once the diagrams are selected, we switch from Compton tensors to their \ac{FxsQED} description, i.e. as in~\eqref{eq:A1mixed}. 
    In particular, we also describe $\gamma^*\gamma^*\gamma\to\pi\pi$ in \ac{FxsQED} (see Table~\ref{tab:ComptonTensors}).
    \item[5)] The selection rules 
    require an adaption of the Yennie-Frautschi-Suura factorisation~\cite{Yennie:1961ad} of the \ac{IR} divergences. 
    More specifically, 
    an additional factor of $1/2$ (shown in purple in the following equation) has to be included on top of the usual expression that is here denoted as $\ieik$, following the convention of~\cite{Engel:2019nfw},
    \begin{align}
    \label{eq:YFSrelation}
    \mathcal{M}^{(1)}\Big|_{q^4_e\, q^4_\pi} \, \supset \quad &2\,\Re\,\beyond{\left(\,
    \begin{gathered}
        \begin{tikzpicture}[scale=0.8]
            \draw (-1.3,0.8) -- (-0.5,0) -- (-1.3,-0.8);
            \draw[photon] (-0.5,0) -- (0.5,0);
            \draw[dashed] (1.3,0.8) -- (0.5,0) -- (1.3,-0.8);
            \draw[photon] (-0.9,0.4) -- (0.2,0.4);
            \fill[gray] (0.5,0.) circle (0.2);
        \end{tikzpicture}
    \end{gathered}
    \,\right)^\dagger
    \cdot
    \begin{gathered}
        \begin{tikzpicture}[scale=0.8]
            \draw (135:1.5) -- (135:0.9) -- (-135:0.9) -- (-135:1.5);
            \draw[dashed] (45:1.5) -- (45:0.9) -- (-45:0.9) -- (-45:1.5);
            \draw[photon] (135:0.9) -- (45:0.9);
            \draw[photon] (-135:0.9) -- (-45:0.9);
            \draw[photon] (0.65,0) -- (1.2,0.);
            \fill[fill=gray]
        (0.6,0) ellipse (0.39 and 0.8);
        \end{tikzpicture}
    \end{gathered}}\Bigg|_\text{IR div.} = \textcolor{purple}{\frac{1}{2}} \, \ieik\Big|_{q_e\, q_\pi} \, \mathcal{M}^{(0)}\Big|_{q^3_e\, q^3_\pi}\,.
\end{align}
This additional factor of $1/2$ indicates again that the selection of the diagrams is only technically motivated. In principle, they should not be split from the rest.
In~\eqref{eq:YFSrelation} we have used the squared amplitudes at tree level 
$\mathcal{M}^{(0)} = \vert \mathcal{A}^{(0)} \vert^2$
and one loop
$\mathcal{M}^{(1)} = 2 \, \text{Re}[ (\mathcal{A}^{(0)})^\dagger \, \mathcal{A}^{(1)} ]$.
Furthermore, as in Figure~\ref{fig:overviewmixed} we have again only shown representative diagrams.
\end{itemize}
With these rules, we estimate the effect of the next-most complicated and much less understood contributions for \eeppg{}, namely those involving $\gamma^*\gamma^*\gamma\to\pi\pi$ rather than only $\gamma^*\gamma^*\to\pi\pi$. For this gauge-invariant and IR finite subset FsQED is a reasonable starting point.
When presenting results later in Section~\ref{sec:results}, we refer to this subset of contributions from the $q_e^4 \,q_\pi^4$ subclass as `beyond'. We will include them in our 'best' prediction, however with the view that they are just indicative of a theoretical error. Hence, whenever they are non-negligible, we will show their size separately.

As can be easily seen from Figure~\ref{fig:e3p5}, the whole subclass $q_e^3 \,q_\pi^5$ is excluded by the rules presented earlier. 
We find the same for \ac{FSC}.
As emphasised earlier in this section, there is currently no physics justification for calculating them in \ac{FsQED}.

\subsection{Comment on other approaches}
In the previous parts of this section, we have discussed how to separate contributions that can be reliably calculated within \ac{FsQED} from those that cannot. 
This separation is based on the discussion in and around Table~\ref{tab:ComptonTensors}, in particular from the requirement to avoid radiative corrections to Compton tensors.
Hence, the argument that prevents us from evaluating the blue contributions in Figure~\ref{fig:overviewmixed} is physical rather than technical.
In fact, (some of) those contributions 
have been calculated in
\ac{FsQED}~\cite{CarloniCalame:2026gdm} and 
in \ac{GVMD}~\cite{PetitRosas:2026iuq,CarloniCalame:2026vfc}.
As in the case of \eepp{}~\cite{Budassi:2024whw}, no large difference between \ac{GVMD} and \ac{FsQED} was observed for \eeppg{} either~\cite{CarloniCalame:2026gdm}.
A potential difference between \mcmule{}'s \ac{FsQED} for $q_e^5 \,q_\pi^3\,+\,$beyond and a calculation of the whole set of corrections can provide important insights into the theory uncertainty of \eeppg{}.
\section{Technical challenges of FsQED}
\label{sec:disperonQED}
When calculating \eeppg{} in \ac{FsQED}, the \ac{FxsQED} description of the $\gamma^* \gamma^* \to \pi\pi$ Compton tensor leads to contributions of the form
\begin{align}
\label{eq:A1mixed}
     \mathcal{A}^{(1)}_{\rm mixed} \supset \begin{gathered}
        \begin{tikzpicture}
            \draw (135:1.5) -- (135:1) -- (-135:1) -- (-135:1.5);
            \draw[dashed] (45:1.5) -- (45:1) -- (-45:1) -- (-45:1.5);
            \draw[photon] (135:1) -- (45:1);
            \draw[photon] (-135:1) -- (-45:1);
            \draw[photon] (-0.7,0) -- (-1.2,0.);
           \fill[fill=gray]
        (0.62,0) ellipse (0.42 and 0.8);
        \end{tikzpicture}
    \end{gathered} \cong \begin{gathered}
        \begin{tikzpicture}
            \draw (135:1.5) -- (135:1) -- (-135:1) -- (-135:1.5);
            \draw[dashed] (45:1.5) -- (45:1) -- (-45:1) -- (-45:1.5);
            \draw[photon] (135:1) -- (45:1);
            \draw[photon] (-135:1) -- (-45:1);
            \draw[photon] (-0.7,0) -- (-1.2,0.);
            \fill[gray] (45:1) circle (0.2);
            \fill[gray] (-45:1) circle (0.2);
        \end{tikzpicture}
    \end{gathered}
    + \begin{gathered}
        \begin{tikzpicture}
            \draw (135:1.5) -- (135:1) -- (-135:1) -- (-135:1.5);
            \draw[dashed] (45:1.5) -- (45:1) -- (-45:1) -- (-45:1.5);
            \draw[photon] (135:1) -- (-45:1);
            \draw[photon] (-135:1) -- (45:1);
            \draw[photon] (-0.7,0) -- (-1.2,0.);
            \fill[gray] (45:1) circle (0.2);
            \fill[gray] (-45:1) circle (0.2);
        \end{tikzpicture}
    \end{gathered}
    + \begin{gathered}
        \begin{tikzpicture}
            \draw (135:1.5) -- (135:1) -- (-135:1) -- (-135:1.5);
            \draw[dashed] (45:1.5) -- (0:0.5) -- (-45:1.5);
            \draw[photon] (135:1) -- (0:0.5);
            \draw[photon] (-135:1) -- (0:0.5);
            \draw[photon] (-0.7,0) -- (-1.2,0.);
            \fill[gray] (0:0.5) circle (0.2);
        \end{tikzpicture}
    \end{gathered}
    \,.
\end{align}
The pion \acp{VFF} entering this description of the hadronic amplitude appear inside the resulting loop integral, preventing its evaluation using standard methods.
This can be circumvented by applying the once-subtracted dispersion relation to every \ac{VFF} that depends on the loop momentum $k^2$
\begin{align}
    \frac{\Fpi(k^2)}{k^2} = \frac{\Fpi(0)}{k^2} - \frac1\pi\dispint \frac{\D \dpone}{\dpone}\frac{\Im \,\Fpi(\dpone)}{k^2-\dpone}\,,
    \label{eq:dispvff}
\end{align}
where $\Fpi(0) =1 $.
Of course, it is always understood that $k^2=k^2+i\delta$.
Rather than evaluating $\Fpi(k^2)$ as a function of the loop momentum, the r.h.s. of~\eqref{eq:dispvff} involves only standard massive photon (which we call disperon) propagators but contains an additional integration over the dispersive parameter $\dpone$ starting from the production threshold $4 m^2_\pi$ up to infinity.
More specifically, we have a split of all diagrams in~\eqref{eq:A1mixed} after applying~\eqref{eq:dispvff}
\begin{align}
\label{eq:AfullFsQED}
    \mathcal{A}^{(1)}_{\rm mixed} \supset
    \underbrace{\pentagon{photon}{photon}{1}{dashed}}_{pp}
    &- \underbrace{
        \frac1\pi\dispint\frac{\Im \,\Fpi(\dpone)}{\dpone}
        \Bigg(
            \pentagon{zigzag, line width=1.2pt}{photon}{1}{dashed} + \,\pentagon{photon}{zigzag, line width=1.2pt}{1}{dashed}
        \Bigg)
    }_{pd}
     \\
    &+\underbrace{
        \frac1{\pi^2}\dispint\dispinttwo
        \frac{\Im \,\Fpi(\dpone)}{\dpone}\frac{\Im \,\Fpi(\dptwo)}{\dptwo}
        \,\pentagon{zigzag,line width=1.2pt}{zigzag,line width=1.2pt}{1}{dashed}
    }_{dd}\,, \nonumber
\end{align}
where the bold zigzag line indicates a disperon.
Diagrams are split into photon-photon ($pp$), photon-disperson ($pd$), and disperon-disperon ($dd$) contributions.

\begin{figure}
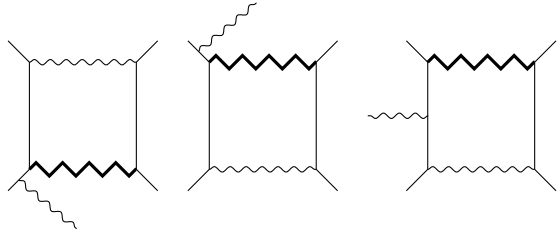

    \centering
    \boxdiagphoton{photon}{zigzag, line width=1.2pt}{2}{}
    \boxdiagphotontwo{zigzag, line width=1.2pt}{photon}{2}{}
    \pentagon{zigzag, line width=1.2pt}{photon}{2}{}
    \caption{
    Topologies contributing to the threshold singularity in \eeppg{}.
    Since the photon can be emitted from either the electron or the pion line, we do not use dashed lines to indicate a pion.
    }
    \label{fig:ee2uug}
\end{figure}

Disperon QED~\cite{Fang:2025mhn} provides a method that allows the consequences of applying~\eqref{eq:dispvff} to be treated efficiently and systematically within a Monte Carlo code such as \mcmule{}.
In particular, there are three challenges that are also visible in~\eqref{eq:AfullFsQED}.
\begin{itemize}
    \item Tedious expressions: 
    The dispersion relation~\eqref{eq:dispvff} creates additional diagrams with massive disperons. 
    Calculating them manually on a case-by-case base is tedious and inefficient.
    Instead, we rely on \OpenLoops{}~\cite{Buccioni:2017yxi,Buccioni:2019sur} with a dedicated disperon QED model~\cite{Fang:2025mhn} for automatised amplitude generation.
    \item Numerical stability and efficiency: The dispersive integral in~\eqref{eq:AfullFsQED} has to be evaluated up to infinity.
    This challenges the numerical stability of the integrand. 
    Hence, instead of using the amplitude provided by \OpenLoops{}
    (in the so-called hybrid precision mode, which computes the bulk of the amplitude in double precision while performing some critical steps in quadruple precision~\cite{Buccioni:2019sur})
    throughout the whole dispersive integration region, above a suitable chosen cut-off value for the dispersive parameter $s_{\rm cut} \gg s, m^2_e, m^2_\pi, \dots$ (for an illustration see~\cite{Kollatzsch:2026ubi}) we switch to an expanded version of the amplitude
    \begin{align}
    \dispint\frac{\Fpi(\dpone)}{\dpone} 
    \, \pentagon{zigzag, line width=1.2pt}{photon}{1}{} 
    &=
    \int_\sth^{s_{\rm cut}} \D\dpone \frac{\Fpi(\dpone)}{\dpone} 
    \, \pentagon{zigzag, line width=1.2pt}{photon}{1}{}\Bigg|_{\text{\OpenLoops{}}} \\
    &+
    \int_{s_{\rm cut}}^\infty \D\dpone \frac{\Fpi(\dpone)}{\dpone} 
    \, \left( \pentagon{zigzag, line width=1.2pt}{photon}{1}{}\Bigg|_{\text{disperon effective theory}} + \mathcal{O}\left(\frac{1}{s_1^3}\right)\right)\,. \notag
    \end{align}
    This expansion, dubbed disperon effective theory in~\cite{Fang:2025mhn}, is based on effective field theory methods and includes terms up to $\mathcal{O}(\{s^2, m^4_e, m^4_\pi, \dots \}/\dponesq)$.
    \item Threshold divergences: The integrand in~\eqref{eq:AfullFsQED} can contain a singularity of the form $(\thrs - \dpone)^{-1}$, where $\thrs$ is the momentum transfer that is available for the pion-pair production and depends on the given diagram.
    For \eeppg{}, topologies that have a $(\thrs - \dpone)^{-1}$ singularity are shown in Figure~\ref{fig:ee2uug}.
    They are part of the $pd$ subset of the amplitude~\eqref{eq:AfullFsQED}.
    Those singularities are dealt with during the integration by applying a subtraction of the whole amplitude at the singular point $s_1 = \thrs$, i.e.
    \begin{align}
    \label{eq:oursubtraction}
    &\dispint\frac{\Fpi(\dpone)}{\dpone} 
    \, \pentagon{zigzag, line width=1.2pt}{photon}{1}{} 
    \\
    &=\dispint\Bigg(\frac{\Fpi(\dpone)}{\dpone}
    \, \pentagon{zigzag, line width=1.2pt}{photon}{1}{}
    -\frac{\Fpi(\thrs)}{\thrs} \, {\rm CT}\Bigg) + \frac{\Fpi(\thrs)}{\thrs}\dispint\ {\rm CT}\,, \notag
    \end{align}
    where the \ac{CT} is given by the singular part of the diagram at $s_1 = \thrs$.
    It can be written as~\cite{Fang:2025mhn}
    \begin{align}
       {\rm CT} = \frac1{\dpone}\Big(\frac{\thrs}{\dpone-\thrs-\io}\Big)^{1+2\epsilon}\,f\,,
    \end{align}
    where $f$ is a universal function that depends on the kinematics of the diagram but not on $\dpone$. 
    Its explicit form can be found in Appendix~\ref{sec:threshold}.
    The first integral in the second line of~\eqref{eq:oursubtraction} is performed numerically within \mcmule{} during the Monte Carlo integration. The second integral over the \ac{CT} is performed analytically~\cite{Fang:2025mhn}.
\end{itemize}

\section{Results}
\label{sec:results}
In this section we present phenomenological results for the scenarios and observables defined during the first phase of the RadioMonteCarLow~2 initiative~\cite{Aliberti:2024fpq}, also using the \ac{VFF} defined there.
For illustration, we show the KLOE-like scenarios (Sections~\ref{sec:KLOE-LA} and~\ref{sec:res-kloe-sa}) also with the pion \ac{VFF} from~\cite{Colangelo:2018mtw}.
All raw data and analysis pipelines can be found at~\cite{McMule:data}
\begin{quote}
    \url{https://mule-tools.gitlab.io/user-library//pion-pair/}
\end{quote}
in the corresponding folders named after the scenarios defined later.
For observables at $\text{N}^{n}\text{LO}$, we introduce the notation
\begin{align}
    \sigma_n=\sigma^{(0)}\,+\,\sigma^{(1)}\,+\,...\,+\sigma^{(n)}\,.
\end{align}
Hence, $\sigma_0=\sigma^{(0)}$.
In \mcmule{}, the \ac{LO} prediction contains all contributions (\ac{ISC}, mixed, and \ac{FSC}).
As discussed in the previous sections, the focus of this work is on the \ac{NLO} contribution
\begin{align}
    \sigma^{(1)} = \sigma^{(1)}_{\rm ISC}\,+\,\sigma^{(1)}_{\rm mixed}\,+\,\sigma^{(1)}_{\rm FSC}\,,
\end{align}
where we include the complete \ac{ISC} ($\sigma^{(1)}_{\rm ISC}$), retain the above discussed subset of the mixed corrections ($\sigma^{(1)}_{\rm mixed}$) calculated using the \ac{FsQED} prescription, and exclude the complete \ac{FSC} ($\sigma^{(1)}_{\rm FSC}$) -- based on our discussions in Section~\ref{sec:compton}.
Thus, the \ac{NLO} corrections included in our results can be written as
\begin{align}\label{eq:dNLOeeppg}
    \sigma^{(1)}_{\rm best\,FsQED} = \sigma_{\rm ISC}^{(1)} + \sigma_{\rm mixed\,FsQED}^{(1)}\big\vert_{q_e^5 \,q_\pi^3}+ \sigma_{\rm mixed\,FsQED}^{(1)}\big\vert_{{\rm beyond}\,(q_e^4\,q_\pi^4)}\,.
\end{align}
To simplify the notation, we combine the mixed corrections to
\begin{align}\label{mix_convention}
    \sigma_{\rm mixed\,FsQED}^{(1)}\big\vert_{q_e^5 \,q_\pi^3\,+\,{\rm beyond}}=\sigma_{\rm mixed\,FsQED}^{(1)}\big\vert_{q_e^5 \,q_\pi^3}+ \sigma_{\rm mixed\,FsQED}^{(1)}\big\vert_{{\rm beyond}\,(q_e^4\,q_\pi^4)}\,,
\end{align}
such that our best prediction includes
\begin{align}
    \label{eq:best}
    \sigma^{(0)}+\sigma^{(1)}_{\rm best\,FsQED} = \sigma_1^{\rm ISC}+\sigma_{\rm mixed\,FsQED}^{(1)}\big\vert_{q_e^5 \,q_\pi^3\,+\,{\rm beyond}}\,,
\end{align}
where we have used
\begin{align}
    \sigma^{\rm ISC}_1 = \sigma^{(0)}+\sigma^{(1)}_{\rm ISC}\,.
\end{align}
To show the size of different contributions w.r.t. \ac{LO} combined with \ac{NLO} \ac{ISC} effects, we define
\begin{align}
    \label{eq:thedelta}
    \delta[X] = \frac{\sigma^{\rm ISC}_1 + X}{\sigma^{\rm ISC}_1} - 1 = \frac{X}{\sigma^{\rm ISC}_1} \,,
\end{align}
which measures the relative impact of the mixed \ac{NLO} correction $X$ with respect to reference observable $\sigma_1^{\rm ISC}$.
For completeness, we also show the size of $\sigma^{\rm ISC}_1$ compared to \ac{LO}
\begin{align}
   \label{eq:NLOISCimpact}
    \quad \hat{\delta}[{\sigma^{(1)}_{\rm ISC}}] = \frac{\sigma^{\rm ISC}_1}{\sigma_0} - 1 
    = \frac{\sigma_{\rm ISC}^{(1)}}{\sigma_0} \,.
\end{align}

\subsection{KLOE-like large-angle scenario}
\label{sec:KLOE-LA}

For the KLOE-like large-angle scenario with $\sqrt{s} = 1.02\,{\rm GeV}$ we define the quantities
\begin{subequations}\label{quantities}
    \begin{align}
        &\theta_{\rm avg}=\frac{\theta^--\theta^++\pi}{2}\,,  & \label{def-thav}\\
        &M_{\pi\pi}^2 = (p_+ + p_-)^2\,, & \label{def-mxx}\\
        &M^2_{\rm trk}=\frac{a}{4}-\frac{\vp_-^2+\vp_+^2}{2}+\frac{(\vp_-^2-\vp_+^2)^2}{4a}\,,&a=(\sqrt{s}-|\vec{p}_-+\vec{p}_+|)^2\,, \label{def-mtrk}
    \end{align}
\end{subequations}
where $\vec{p}_{\pm}$, $p_{\pm}$, and $\theta^\pm$ denote the three-momenta, four-momenta, and polar angles of the  $\pi^\pm$ and $\vp_{\pm} = |\vec{p}_\pm|$.
The track mass in~\eqref{def-mtrk} is defined as the solution of
\begin{align}
    \left(\sqrt{s}-\sqrt{|\vp_+|^2+M_{\rm trk}^2}-\sqrt{|\vp_-|^2+M_{\rm trk}^2}\right)^2-\left|\vec{p}_++\vec{p}_-\right|^2\equiv 0 \,,
\end{align}
which is used in the experiment to distinguish between the various final states in $ee\to XX\gamma$.
We impose the following cuts
\begin{subequations}\label{mcc:cutKLOE-la}
\begin{align}
&E_\gamma > 20\,{\rm MeV}& &\mbox{and}& 
&50^\circ\le\theta_\gamma\le 130^\circ\, , & \label{mcc:cutKLOE-la-y} 
\\
&|\vp^z_\pm| > 90\,{\rm MeV} \  \mbox{or}\  \vp^\perp_\pm > 160\,{\rm MeV}&
&\mbox{and}& &50^\circ\le\theta^\pm\le 130^\circ\, , & \label{eq:kloe-la-th} 
\\
&0.1\,{\rm GeV}^2 \le M_{\pi\pi}^2 \le 0.85\,{\rm GeV}^2\, , &\label{cut:KLOE-LA-mass} 
\end{align}
\end{subequations}
where $\vp^z_\pm$ and $\vp^\perp_\pm$ denote the $z$ and transverse components of the $\pi^{\pm}$ and~\eqref{mcc:cutKLOE-la-y} is to be understood as requiring at least one such photon. 
If further the cut
\begin{align}
 130\,{\rm MeV} \le M_{\rm trk} \le 220\,{\rm MeV}\,, \label{KLOE-LA-mxxc}
\end{align}
is applied to the distribution of $M_{\pi\pi}$, we obtain the observable $M_{\pi\pi}^c$.

\begin{figure}
    \centering
    \subfloat[
        Distribution for $\thavg$ defined in~\eqref{def-thav}.
        ]{\includegraphics[width=0.8\textwidth]{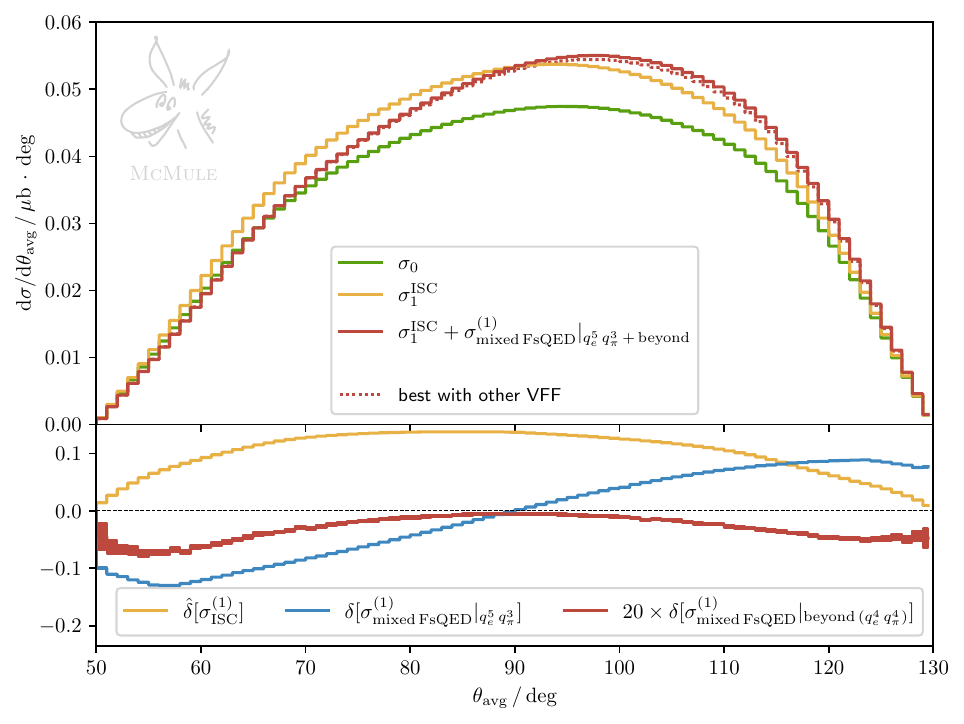}\label{fig:kloe-la-thav}}\\
    \subfloat[
    Distribution for $\theta^+$.
    ]{\includegraphics[width=0.8\textwidth]{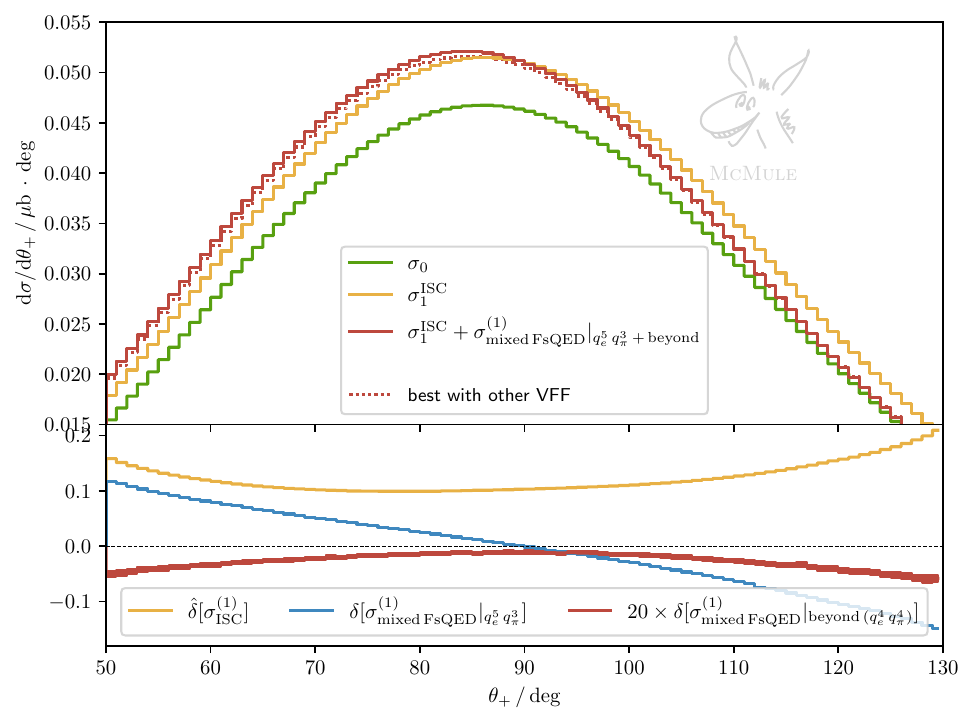}\label{fig:kloe-la-th}}
    \caption{
    Results for the KLOE-like large-angle scenario.
    The top panels show distributions for various contributions.
    The bottom panels show some of the $\delta$'s defined above. 
    The dotted curve in the top panels shows the best prediction~\eqref{eq:best} using the \ac{VFF} of~\cite{Colangelo:2018mtw}.
    }
    \label{fig:kloe-la-ppg}
\end{figure}

\begin{figure}
    \centering
    \subfloat[Distribution for $M_{\pi\pi}$ defined in~\eqref{def-mxx}.]{\includegraphics[width=0.8\textwidth]{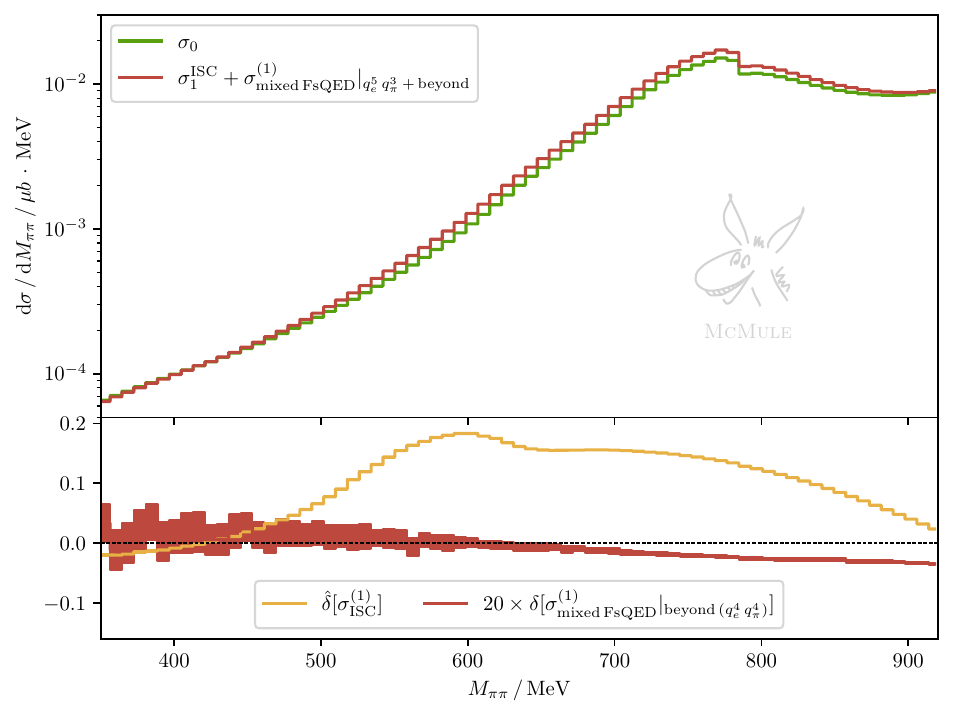}\label{fig:kloe-la-lmxx}}\\
    \subfloat[Distribution for $M_{\pi\pi}^c$, i.e. $M_{\pi\pi}$  with the additional cut~\eqref{KLOE-LA-mxxc}.]{\includegraphics[width=0.8\textwidth]{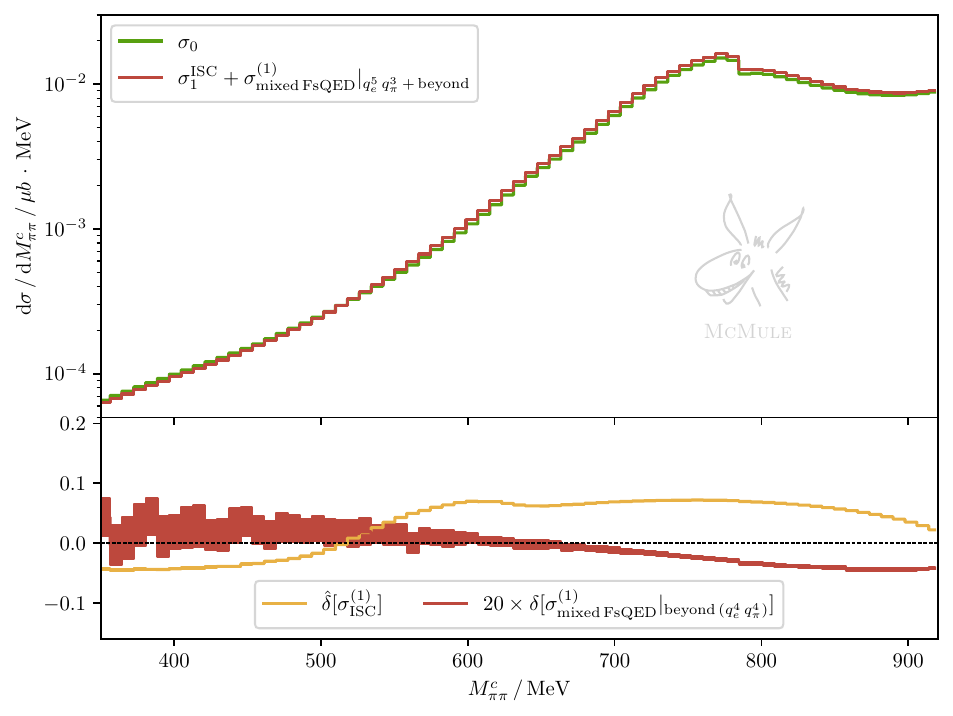}\label{fig:kloe-la-lmxxc}}
    \caption{
    Results for the KLOE-like large-angle scenario (continued).
    The top panels show distributions for various contributions.
    The bottom panels show some of the $\delta$'s defined above.
    We do not show the effect of $\sigma_{\rm mixed\,FsQED}^{(1)}\big\vert_{q_e^5 \,q_\pi^3}$ in the bottom panels as it is zero.
    Given the smallness of the remaining $\sigma_{\rm mixed\,FsQED}^{(1)}\big\vert_{{\rm beyond}\,(q_e^4\,q_\pi^4)}$ (see Table~\ref{tab:impactbeyond}), the red curve in the top panels overlaps with $\sigma_1^{\rm ISC}$ (not shown).
    }
    \label{fig:kloe-la-ppg2}
\end{figure}

In Figures~\ref{fig:kloe-la-ppg} and~\ref{fig:kloe-la-ppg2}, we show distributions for the KLOE-like large-angle scenario for $\thavg$ (Figure~\ref{fig:kloe-la-thav}), $\theta^+$ (Figure~\ref{fig:kloe-la-th}), $M_{\pi\pi}$ (Figure~\ref{fig:kloe-la-lmxx}), and $M_{\pi\pi}^c$ (Figure~\ref{fig:kloe-la-lmxxc}).

In Figure~\ref{fig:kloe-la-ppg} we focus on observables that are $C$-odd and hence, obtain corrections from $\sigma_{\rm mixed\,FsQED}^{(1)}\big\vert_{q_e^5 \,q_\pi^3}$.
As clearly visible, the \ac{NLO} \ac{ISC} are symmetric and their effect reaches up to $15-20\%$ compared to the full \ac{LO}.
The \ac{NLO} mixed corrections also change the result up to around $15\%$, but
in addition, they clearly modify the shape of the distribution.
Hence, for this KLOE scenario, mixed corrections
cannot be neglected.
In fact, they must be included in the best possible way. 
It is reassuring to see, however, that
the huge impact stems from the comparatively better controlled contribution
$\sigma_{\rm mixed\,FsQED}^{(1)}\big\vert_{q_e^5 \,q_\pi^3}$ 
and not from $\sigma_{\rm mixed\,FsQED}^{(1)}\big\vert_{{\rm beyond}\,(q_e^4\,q_\pi^4)}$.
The latter, shown relative to the \ac{NLO} \ac{ISC}, has to be multiplied by a factor of $\mathcal{O}(10)$ in order for its shape to become visible. 
The effect is symmetric. We use  $\sigma_{\rm mixed\,FsQED}^{(1)}\big\vert_{{\rm beyond}\,(q_e^4\,q_\pi^4)}$ as an estimate of the size of corrections ($q_\pi^4 q_e^4$ and higher powers in $q_\pi$) that cannot be reliably calculated within \ac{FsQED}. Unlike in the other scenarios considered (see Table~\ref{tab:impactbeyond}), for the KLOE-like large-angle scenario they have a non-negligible impact and reach the $0.1\%$ level. 
This limits the achievable theoretical precision to the sub-percent level in this scenario.

\begin{table}[]
\centering
\begin{tabular}{p{4.5cm} p{2.8cm} p{2.5cm} p{2.2cm}p{1.9cm}}
\toprule
 & KLOE-like \newline large-angle    & KLOE-like \newline small-angle    & BESIII-like & B           \\
 \midrule
$
\delta[\sigma_{\rm mixed\,FsQED}^{(1)}\big\vert_{{\rm beyond}\,(q_e^4\,q_\pi^4)}]$  & $-1.122(3)\cdot 10^{-3}$ & $-2.66(1)\cdot 10^{-5}$ & $-1.4(3)\cdot 10^{-6}$&$-3(4)\cdot 10^{-8}$
 \\
\bottomrule
\end{tabular}
\caption{
Impact of $\sigma_{\rm mixed\,FsQED}^{(1)}\big\vert_{{\rm beyond}\,(q_e^4\,q_\pi^4)}$ for the total cross section for different scenarios.}
\label{tab:impactbeyond}
\end{table}

Figures~\ref{fig:kloe-la-lmxx} and~\ref{fig:kloe-la-lmxxc} show the invariant mass of the pion pair, without and with the additional cut on the track mass~\eqref{KLOE-LA-mxxc}, respectively. 
At \ac{LO}, the reconstructed track mass satisfies $M_{\rm trk}=m_\pi$, which lies within the imposed track-mass window~\eqref{KLOE-LA-mxxc}. 
Therefore, the \ac{LO} distributions in Figures~\ref{fig:kloe-la-lmxx} and~\ref{fig:kloe-la-lmxxc} are identical. 
At \ac{NLO}, however, real emission is restricted and certain events are removed with the cut~\eqref{KLOE-LA-mxxc}.
This suppression is clearly visible for the \ac{NLO} \ac{ISC} when comparing the two figures in Figures~\ref{fig:kloe-la-ppg2}.
For the mixed corrections, this effect is smaller.
In Figure~\ref{fig:kloe-la-ppg2}, the $\rho$-$\omega$ mixing at $m_\omega \simeq 782\,{\rm MeV}$ is visible.
This resonance is part of the \ac{VFF}.

The red dotted line in the upper panels of Figure~\ref{fig:kloe-la-ppg} shows the best prediction~\eqref{eq:best} with the pion \ac{VFF} from~\cite{Colangelo:2018mtw}. In this case, the vacuum-polarisation contributions are not included in the \ac{VFF} and have to be explicitly added.
Taking this into account, the difference to this prediction with the \ac{VFF} from~\cite{Aliberti:2024fpq} (red solid line) is around $1\,\%$, highlighting the impact of the \ac{VFF} on the absolute cross section.
Its impact on the relative size of the higher order corrections is negligible.
The implementation in \mcmule{} can deal with any suitable \ac{VFF}.

\subsection{KLOE-like small-angle scenario}
\label{sec:res-kloe-sa}
\begin{figure}
    \centering
    \subfloat[
        Distribution for $\thavg$ defined in~\eqref{def-thav}.
        ]{\includegraphics[width=0.8\textwidth]{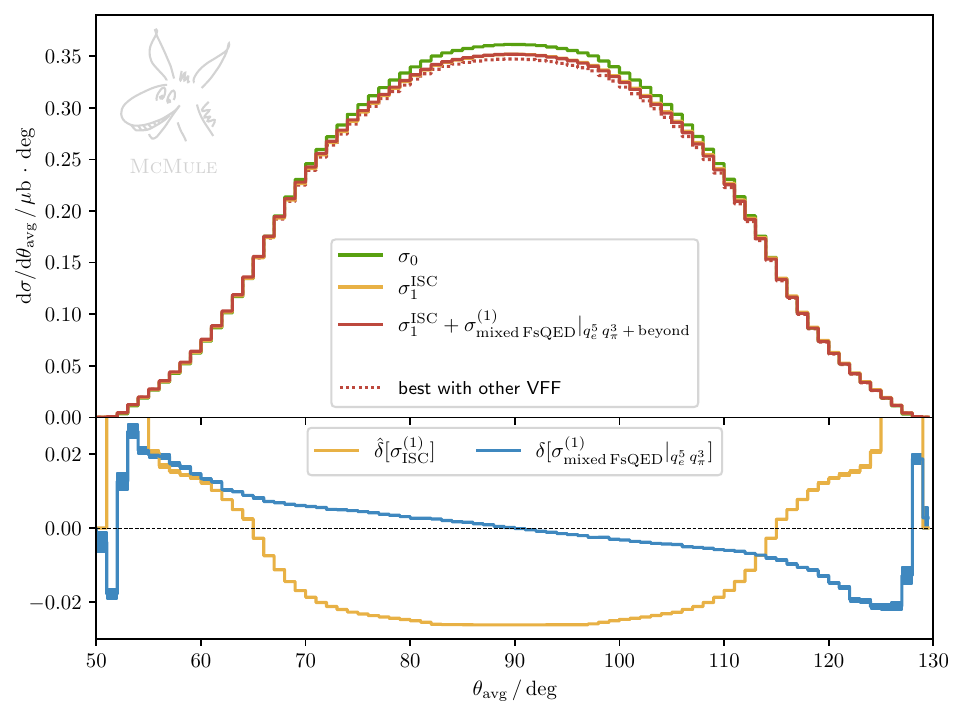}\label{fig:kloe-sa-thav}}\\
    \subfloat[
    Distribution for $\theta^+$.
    ]{\includegraphics[width=0.8\textwidth]{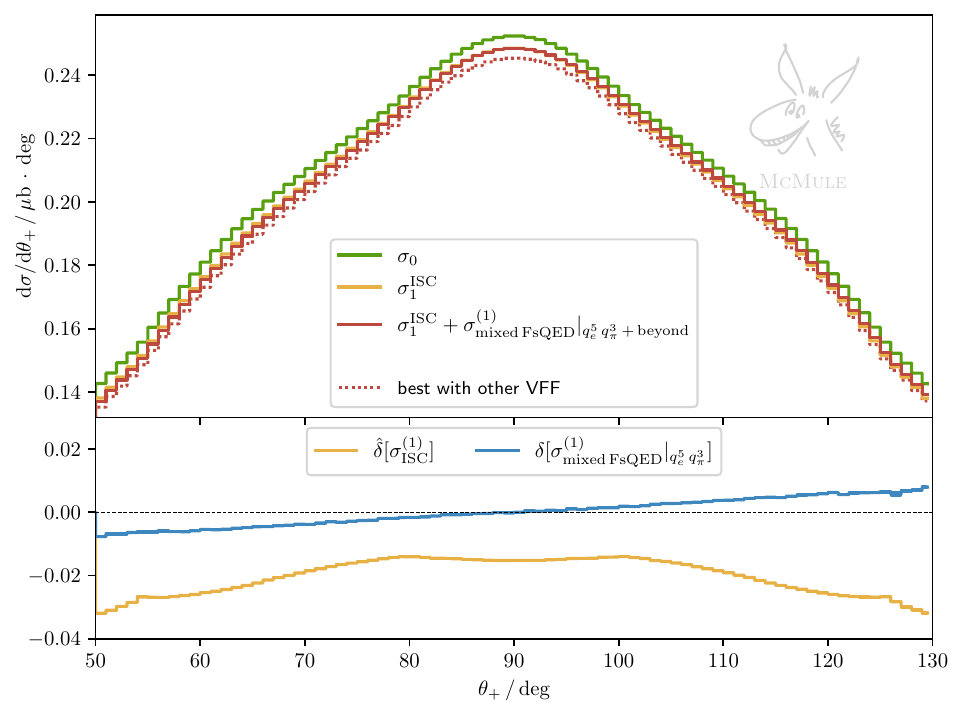}\label{fig:kloe-sa-th}}
    \caption{
    Results for the KLOE-like small-angle scenario.
    The top panels show distributions for various contributions.
     Given the smallness of $\sigma_{\rm mixed\,FsQED}^{(1)}\big\vert_{\rm beyond}$ (see Table~\ref{tab:impactbeyond}), the red curve overlaps the blue one.
     The dotted curve in the top panels shows the best prediction~\eqref{eq:best} using the
     \ac{VFF} of~\cite{Colangelo:2018mtw}.
    The bottom panels show some of the $\delta$'s defined above.
    The tiny effect of $\sigma_{\rm mixed\,FsQED}^{(1)}\big\vert_{{\rm beyond\,(q_e^4\,q_\pi^4)}}$ is not shown explicitly in the lower panels.
    }
    \label{fig:kloe-sa-ppg}
\end{figure}

Also the KLOE-like small-angle has $\sqrt{s}=1.02\,{\rm GeV}$.
However, the photon is not directly detected. 
The angle $\theta_{\widetilde\gamma}$ associated with the `untagged' photon momentum $\vp_{\widetilde\gamma}\equiv -(\vec{p}_{+}+\vec{p}_{-})$ is assumed to be small w.r.t.\ the beam.
More precisely, the cuts we apply are
\begin{subequations}\label{eq:kloesa-cuts}
\begin{align}
&\theta_{\widetilde\gamma}\le 15^\circ \quad \mbox{or} \quad \theta_{\widetilde\gamma}>165^\circ
\, , \\
&|\vp^z_\pm| > 90\,{\rm MeV} \  \mbox{or}\  \vp^\perp_\pm > 160\,{\rm MeV}&
&\mbox{and}& &50^\circ\le\theta^\pm\le 130^\circ\, , & 
\\
&0.35\,{\rm GeV}^2 \le M_{\pi\pi}^2 \le 0.95\,{\rm GeV}^2\, . &
\end{align}
\end{subequations}
In Figure~\ref{fig:kloe-sa-ppg}, we show results for  $\thavg$ (Figure~\ref{fig:kloe-sa-thav}) and  $\theta^+$ (Figure~\ref{fig:kloe-sa-th}).
Given their small size (see Table~\ref{tab:impactbeyond}), we no longer show the impact of
$\sigma_{\rm mixed\,FsQED}^{(1)}\big\vert_{{\rm beyond}\,(q_e^4\,q_\pi^4) }$
explicitly.
Although
$\sigma_{\rm mixed\,FsQED}^{(1)}\big\vert_{{\rm beyond}\,(q_e^4\,q_\pi^4)}$
is only an estimate of the more complicated $q_\pi^4 q_e^4$ contributions, its general smallness provides a useful indication that these contributions are unlikely to be overly large.
The impact of both the \ac{NLO} \ac{ISC} and $\sigma_{\rm mixed,FsQED}^{(1)}\big\vert_{q_e^5 q_\pi^3}$ are significantly smaller than in the KLOE-like large-angle scenario shown in Section~\ref{sec:KLOE-LA}.
They reach $3\%$ w.r.t. \ac{LO} for \ac{NLO} \ac{ISC} and $1\%$ in the bulk of the phase space w.r.t. $\sigma_1^{\rm ISC}$ for $\sigma_{\rm mixed,FsQED}^{(1)}\big\vert_{q_e^5 q_\pi^3}$. 
Again, we clearly see the symmetric and asymmetric behaviour of the contributions.
Furthermore, as for the KLOE-like large-angle scenario in Section~\ref{sec:KLOE-LA}, the impact of using a different \ac{VFF} is noticeable for the absolute cross section, but very small for the shape of the distributions.

\subsection{BESIII-like scenario}
In this scenario we increase the energy to $\sqrt{s}=4\,\rm{GeV}$. 
In order to detect a charged pion or photon, they have to pass the selection cuts 
\begin{subequations}\label{eq:bescut}
\begin{align}\label{mcc:bescut}
&|\cos \theta^{\pm}|<0.93& &\mbox{and}&  &\vp_{\pm}^\perp >300\,{\rm MeV}\, ,& \\
& \big(|\cos\theta_\gamma| < 0.8 \ \mbox{and}\ E_\gamma > 25\,{\rm MeV}\big)&  &\mbox{or}& 
&\big(0.86 < |\cos \theta_\gamma| < 0.92\ \mbox{and} \ E_\gamma > 50\,{\rm MeV}\big)\,,
\end{align}
\end{subequations}
where the cuts on the photon cover regions both close to the beam axis and at larger angles.
In addition, we require that precisely one such photon has $E_\gamma \ge 400\,{\rm MeV}$.

\begin{figure}
    \centering
    \subfloat[
        Distribution for $\thavg$ for the  BESIII-like scenario.
    ]{
        \includegraphics[width=0.8\textwidth]{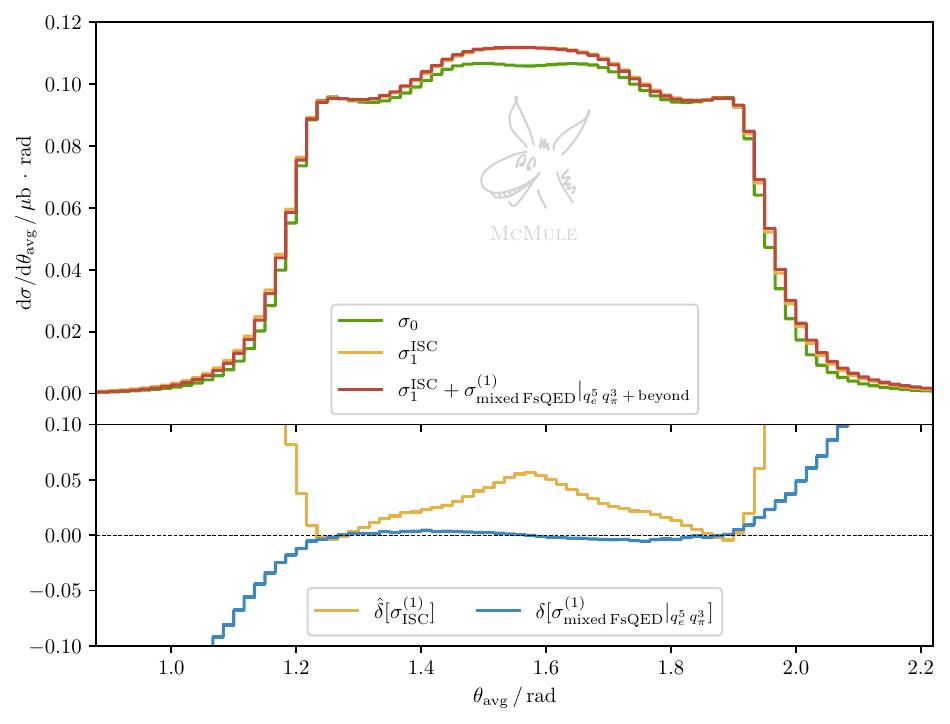}
        \label{fig:bes-ppg}
    }\\
    \subfloat[
        Distribution for $\thavg$ for the $B$ scenario.
    ]{
        \includegraphics[width=0.8\textwidth]{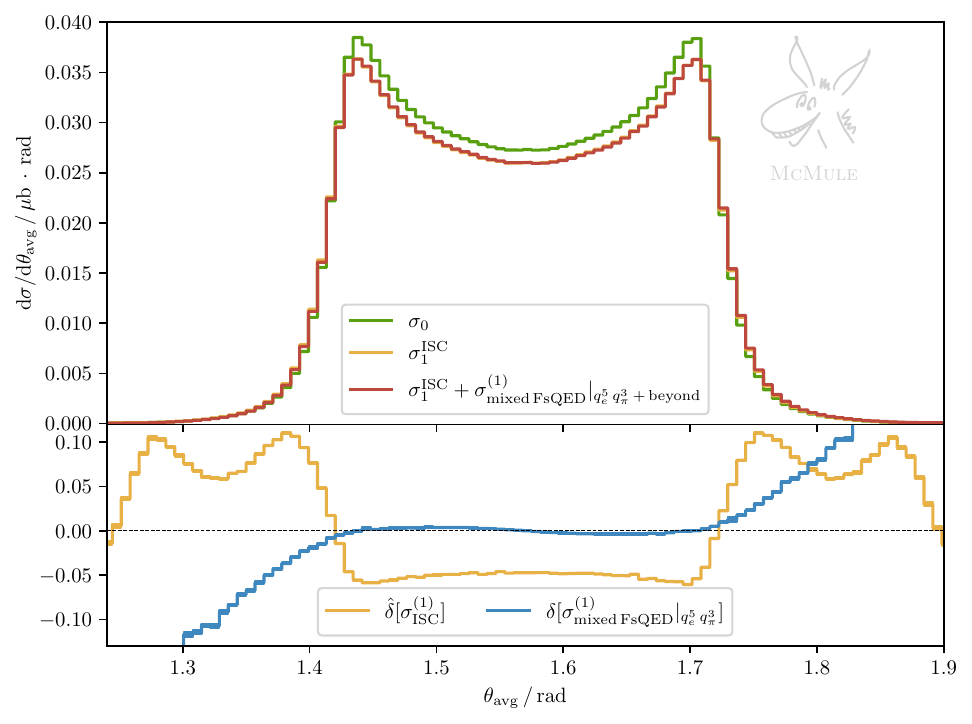}
        \label{fig:b-ppg}
    }
    \caption{
        Results for $\thavg$ defined in~\eqref{def-thav} for the BESIII-like and $B$ scenarios.
        The top panels show distributions for various contributions.
        Given the smallness of the mixed corrections,
        the red curve almost completely overlaps the yellow one.
        The bottom panels show some of the $\delta$'s defined above.
        The tiny effect of $\sigma_{\rm mixed\,FsQED}^{(1)}\big\vert_{{\rm beyond\,(q_e^4\,q_\pi^4)}}$ (see Table~\ref{tab:impactbeyond}) is not shown explicitly in the lower panels.
    }
\end{figure}

In Figure~\ref{fig:bes-ppg}, we show the results for $\thavg$. 
Except at the edges of the distribution, the impact of the \ac{NLO} \ac{ISC} reaches its maximum around $\thavg=\pi/2$, where it amounts to about $5\%$ w.r.t. the \ac{LO} result.
$\sigma_{\rm mixed,FsQED}^{(1)}\big\vert_{q_e^5 ,q_\pi^3}$ is relatively small over most of the phase space, retaining its characteristic asymmetry.
Towards the edges of the distribution, the ratios in the second panel become large, mainly because the \ac{LO} differential cross section approaches zero in these regions.
As can be seen from Table~\ref{tab:impactbeyond}, the impact of $\sigma_{\rm mixed\,FsQED}^{(1)}\big\vert_{{\rm beyond}\,(q_e^4\,q_\pi^4) }$ is negligible.

The significant reduction of the impact of mixed corrections (and consequently also \ac{FSC}) at higher $\sqrt{s}$ is mostly due to the pion \ac{VFF} suppression following from $F_\pi^V(q^2) \to 0$ for $q^2 \to \infty$.
Additionally, the \ac{ISC} are enhanced due to large collinear logarithms. 
This has already been observed in~\cite{Aliberti:2024fpq}.

\subsection{\texorpdfstring{\boldmath $B$ scenario}{B scenario}}
This scenario is inspired by $B$ factories
with an energy of $\sqrt{s}=10\,{\rm GeV}$.
We impose the cuts
\begin{subequations}\label{eq:Bcut}
\begin{align}
&0.65\,{\rm rad}\le\theta^\pm\le 2.75\,{\rm rad}&  &\mbox{and}& &\vp_\pm > 1\,{\rm GeV}\, ,& \\
&0.6\,{\rm rad}\le\theta_\gamma\le 2.7\,{\rm rad}&  &\mbox{and}& &E_\gamma>3\,{\rm GeV}\, ,& \label{eq:Bcuty}
\end{align}
where~\eqref{eq:Bcuty} is to be understood as requiring at least one such photon.
We denote the most energetic photon passing the cut by $\gamma^{(h)}$.
Furthermore, introducing $M_{\pi\pi\gamma}^2\equiv(p_+ + p_- + p_{\gamma^{(h)}})^2$ we require
\begin{align}
&\theta_{\gamma^{(h)},\widetilde\gamma} = 
\sphericalangle(\vec{p}_{\gamma^{(h)}},\vec{p}_{\widetilde\gamma}) < 0.3\,{\rm rad}&  
&\mbox{and}& &M_{\pi\pi\gamma} > 8\,{\rm GeV}\, .& \label{eq:Bcuth}
\end{align}
\end{subequations}
The second cut in \eqref{eq:Bcuth} is to suppress secondary photons.

In Figure~\ref{fig:b-ppg}, we show again the result for $\thavg$.
The \ac{NLO} \ac{ISC} reach about $5\%$ w.r.t. \ac{LO} in the bulk of the phase space.
Given the high $\sqrt{s}$, mixed corrections (and \ac{FSC}) become even less significant than in the KLOE-like and BESIII-like scenarios.
\section{Conclusion and outlook}
\label{sec:conclusions}
Triggered by the ongoing RadioMonteCarLow~2 effort, several Monte Carlo codes are actively working towards improved predictions for $ee\to\pi\pi\gamma$.
In this work, we have presented results for \ac{NLO} corrections that can be reliably calculated within \ac{FsQED}.
Beside the previously known \ac{ISC}, this is the $q_e^5 \,q_\pi^3$ subset of mixed corrections.
While we find a negligible impact of them for scenarios at higher $\sqrt{s}$, they can reach several precent w.r.t. $\sigma^{\rm ISC}_1$ for the KLOE-like scenarios.
The KLOE-like large-angle scenario is most affected.
We emphasise that this subset only contains the Compton tensor $\gamma^*\gamma^*\to \pi\pi$.
The more complicated and less well-understood $\gamma^*\gamma^*\gamma\to\pi\pi$ Compton tensor has to be evaluated together with radiative corrections to the Compton tensors, for which a physically motivated description is even less developed.

There are two avenues of refining the presented \eeppg{} prediction, one dominated by technical aspects, the other by conceptual  non-perturbative aspects. 

On the technical side, it is possible to advance the corrections associated to the electron line.
These are in particular \ac{ISC} effects that are already available beyond \ac{NLO} matched to a parton shower~\cite{Budassi:2026lmr}. 
A fixed order calculation of the \ac{ISC} effects at \ac{NNLO} is planned for the near future from \mcmule{}~\cite{Kollatzsch:2026TI,Fang:2026dbs}.
Furthermore, in principle also the subset of mixed corrections ($q_e^{5} \,q_\pi^3$) which are well defined in \ac{FsQED} could be extended to two loops ($q_e^{5+2} \,q_\pi^3$). Finally, the methods presented here can easily be adapted to kaon pair production.

On the conceptual side, progress could be made by using a more sophisticated description of the Compton tensor.
Strictly speaking, also effects that go beyond the pion pole approximation for the Compton tensor (\ac{FxsQED}) have to be considered.
While these effects in $\gamma^* \gamma^* \to \pi \pi$ are expected to be small, for $\gamma^* \gamma^* \gamma \to \pi \pi$, their size in \eeppg{} could be estimated in a first step by approximating them with $\gamma^* \gamma^* \gamma \to \rho(770) \to \pi \pi$, where the resonance $\rho(770)$ is produced as an (on-shell) intermediate state~\cite{Hoferichter:2026rescattering,Hoferichter:note2026}.
A first study~\cite{CarloniCalame:2026gdm} suggest a negligible effect. 
However, a reliable computation of final-state emission is currently not in sight.

Hence, our aim was to provide a physics-motivated subset of corrections that can be computed reliably. A comparison with a calculation of the full set of corrections, including effects that cannot be reliably calculated in \ac{FsQED}, will provide a lower estimate of the theory uncertainty associated to \eeppg{}.
First indications presented here suggest a noticeable impact for the KLOE-like large-angle scenario.
A detailed comparison is planned within the RadioMonteCarLow~2 effort.

\subsection*{Acknowledgement}
We wish to thank Martin~Hoferichter and Peter~Stoffer for discussions and clarifications regarding the description of pions and pion \acp{VFF}.
We also wish to thank the RadioMonteCarLow~2 community for enlightening discussions about different contributions and conventions in $ee \to \pi\pi \gamma$ in \href{https://indico.global/event/16381/}{Torino}.
It is a pleasure to thank Fedor~Ignatov and Graziano~Venanzoni for discussions and clarification about the required theoretical precision for the upcoming KLOE-NXT analysis~\cite{KLOE:private}.
We thank Yannick~Ulrich and Marco~Rocco for their support with technical aspects related to \mcmule{}.

YF and SK acknowledge support by the Swiss National Science Foundation (SNSF) under grants 10006722 and 207386, respectively.
The research of MZ was supported by the SNSF under the contract TMSGI2-211209.

\appendix
\section{Threshold counterterm}
\label{sec:threshold}
In~\cite{Fang:2025mhn}, we have presented the expression for $f$ that was needed for \eepp{}.
In this appendix, we extend this definition to a more general case that can be applied to \eeppg{}.
Since pentagon diagrams can be reduced to box diagrams via Cayley reduction~\cite{Denner:2002ii,Bern:1992em,Bern:1993kr}, we consider a generic box integral using the notation inspired by~\cite{Ellis:2007qk} but keep our momentum convention, i.e.
\begin{align}
    I(\tilde{p}^2_1,\tilde{p}^2_2,\tilde{p}^2_3,\tilde{p}^2_4; \tilde{s}_{12},\tilde{s}_{23},m^2_1,m^2_2,m^2_3,m^2_4) = 
    \begin{gathered}
        \begin{tikzpicture}[scale=0.5]
            \draw (-1,1) -- (-2,2) node[left] {$\tilde{p}_1$};
            \draw (-1,-1) -- (-2,-2) node[left] {$\tilde{p}_2$};
            \draw (-1,1) -- (1,1) node[midway, above] {$m_3$};
            \draw (-1,1) -- (-1,-1) node[midway, left] {$m_2$};
            \draw (1,-1) -- (1,1) node[midway, right] {$m_4$};
            \draw (-1,-1) -- (1,-1) node[midway, below] {$m_1$};
            \draw (1,1) -- (2,2) node[right] {$\tilde{p}_3$};
            \draw (1,-1) -- (2,-2) node[right] {$\tilde{p}_4$};
        \end{tikzpicture}
    \end{gathered}
    \,,
\end{align}
where $\tilde{p}_1 + \tilde{p}_2 = \tilde{p}_3 + \tilde{p}_4$, $\tilde{s}_{12} = (\tilde{p}_1 + \tilde{p}_2)^2$ and $\tilde{s}_{23} = (\tilde{p}_2 - \tilde{p}_3)^2$.
In the case of $ee\to\pi\pi\gamma$ with $m_3 = \sqrt{\dpone}$ and $m_1 = 0$, the photon is emitted from or assigned to the $\tilde{p}_1$ or $\tilde{p}_3$ leg or crossed versions of it and $\thrs = \tilde{s}_{12}$.
Hence, the general integral simplifies to
\begin{align}
    &I(m^2_e,\tilde{p}^2_2,\tilde{p}^2_3,m^2_\pi; \tilde{s}_{12} = \thrs,\tilde{s}_{23},0,m^2_e,\dpone,m^2_\pi)  \\
    &= \frac1{\dpone}\Big(\frac{\thrs}{\dpone-\thrs-\io}\Big)^{1+2\epsilon}f(\tilde{p}^2_2,\tilde{p}^2_3,\thrs,\tilde{s}_{23},m_e,m_\pi) +\mathcal{O}(\dpone-\thrs)^0 = \rm{CT}\,, \notag
\end{align}
where the function $f$ is defined as
\begin{align}
\label{eq:fexplicit}
f = & \left(\frac{\mu^2}{\thrs}\right)^\epsilon 
\frac{\chi_t}{m_e m_\pi(\chi_t^2 - 1)} 
\Bigg\{
\log(-\chi_t)\left(\frac{1}{\epsilon} - 2\log(1-\chi_t^2)\right)
\nonumber\\
& + \bigg[
\zeta_2 - \log^2(\chi_e) - \log^2(\chi_\pi) - \mathrm{Li}_2(\chi_t^2)
\nonumber\\
&\quad
- \mathrm{Li}_2(1+\chi_e \chi_\pi \chi_t)
- \mathrm{Li}_2\!\left(1+\frac{\chi_t \chi_\pi}{\chi_e}\right)
- \mathrm{Li}_2\!\left(1+\frac{\chi_t}{\chi_e \chi_\pi}\right)
- \mathrm{Li}_2\!\left(1+\frac{\chi_t \chi_e}{\chi_\pi}\right)
\bigg]
\Bigg\}\,,
\end{align}
together with the $\chi$
\begin{align}
\chi_e^2 = \frac{1-\sqrt{1-\frac{4m_e^2\,\thrs}{(m_e^2-\tilde{p}_2^2+\thrs)^2}}}{1+\sqrt{1-\frac{4m_e^2\,\thrs}{(m_e^2-\tilde{p}_2^2+\thrs)^2}}}\,,\quad
\chi_\pi^2 = \frac{1-\sqrt{1-\frac{4m_\pi^2\,\thrs}{(m_\pi^2-\tilde{p}_3^2+\thrs)^2}}}{1+\sqrt{1-\frac{4m_\pi^2\,\thrs}{(m_\pi^2-\tilde{p}_3^2+\thrs)^2}}}\,,\quad
\chi_t = \frac{1-\sqrt{\frac{(m_e+m_\pi)^2-\tilde{s}_{23}}{(m_e-m_\pi)^2-\tilde{s}_{23}}}}{1+\sqrt{\frac{(m_e+m_\pi)^2-\tilde{s}_{23}}{(m_e-m_\pi)^2-\tilde{s}_{23}}}}\,.
\end{align}

\bibliographystyle{JHEP}
\bibliography{pions}

\providecommand{\href}[2]{#2}\begingroup\raggedright\begin{thebibliography}{10}

\bibitem{Aliberti:2025beg}
R.~Aliberti et~al., \emph{{The anomalous magnetic moment of the muon in the
  Standard Model: an update}},
  \href{https://doi.org/10.1016/j.physrep.2025.08.002}{\emph{Phys. Rept.}
  {\bfseries 1143} (2025) 1}
  [\href{https://arxiv.org/abs/2505.21476}{{\ttfamily 2505.21476}}].

\bibitem{Borsanyi:2020mff}
S.~Borsanyi et~al., \emph{{Leading hadronic contribution to the muon magnetic
  moment from lattice QCD}},
  \href{https://doi.org/10.1038/s41586-021-03418-1}{\emph{Nature} {\bfseries
  593} (2021) 51} [\href{https://arxiv.org/abs/2002.12347}{{\ttfamily
  2002.12347}}].

\bibitem{RBC:2023pvn}
{\scshape RBC, UKQCD} collaboration, T.~Blum et~al., \emph{{Update of Euclidean
  windows of the hadronic vacuum polarization}},
  \href{https://doi.org/10.1103/PhysRevD.108.054507}{\emph{Phys. Rev. D}
  {\bfseries 108} (2023) 054507}
  [\href{https://arxiv.org/abs/2301.08696}{{\ttfamily 2301.08696}}].

\bibitem{Djukanovic:2024cmq}
D.~Djukanovic, G.~von Hippel, S.~Kuberski, H.~B. Meyer, N.~Miller, K.~Ottnad
  et~al., \emph{{The hadronic vacuum polarization contribution to the muon
  $g-2$ at long distances}},
  \href{https://doi.org/10.1007/JHEP04(2025)098}{\emph{JHEP} {\bfseries 04}
  (2025) 098} [\href{https://arxiv.org/abs/2411.07969}{{\ttfamily
  2411.07969}}].

\bibitem{Muong-2:2025xyk}
{\scshape Muon $g-2$} collaboration, D.~P. Aguillard et~al., \emph{{Measurement
  of the Positive Muon Anomalous Magnetic Moment to 127~ppb}},
  \href{https://doi.org/10.1103/7clf-sm2v}{\emph{Phys. Rev. Lett.} {\bfseries
  135} (2025) 101802} [\href{https://arxiv.org/abs/2506.03069}{{\ttfamily
  2506.03069}}].

\bibitem{Aliberti:2024fpq}
R.~Aliberti et~al., \emph{{Radiative corrections and Monte Carlo tools for
  low-energy hadronic cross sections in $e^+ e^-$ collisions}},
  \href{https://doi.org/10.21468/SciPostPhysCommRep.9}{\emph{SciPost Phys.
  Comm. Rep.} {\bfseries 2025} (2025) 9}
  [\href{https://arxiv.org/abs/2410.22882}{{\ttfamily 2410.22882}}].

\bibitem{Budassi:2026lmr}
E.~Budassi, C.~M. Carloni~Calame, M.~Ghilardi, A.~Gurgone, G.~Montagna,
  M.~Moretti et~al., \emph{{Radiative return at NLOPS accuracy}},
  \href{https://doi.org/10.1007/JHEP05(2026)221}{\emph{JHEP} {\bfseries 05}
  (2026) 221} [\href{https://arxiv.org/abs/2601.19530}{{\ttfamily
  2601.19530}}].

\bibitem{Fadin:2023phc}
V.~S. Fadin and R.~N. Lee, \emph{{Two-loop radiative corrections to $e^+ e^-
  \to \gamma\gamma^{*}$ cross section}},
  \href{https://doi.org/10.1007/JHEP11(2023)148}{\emph{JHEP} {\bfseries 11}
  (2023) 148} [\href{https://arxiv.org/abs/2308.09479}{{\ttfamily
  2308.09479}}].

\bibitem{Badger:2023xtl}
S.~Badger, J.~Kry{\'s}, R.~Moodie and S.~Zoia, \emph{{Lepton-pair scattering
  with an off-shell and an on-shell photon at two loops in massless QED}},
  \href{https://doi.org/10.1007/JHEP11(2023)041}{\emph{JHEP} {\bfseries 11}
  (2023) 041} [\href{https://arxiv.org/abs/2307.03098}{{\ttfamily
  2307.03098}}].

\bibitem{PetitRosas:2025xhm}
P.~Petit~Ros{\`a}s and W.~J. Torres~Bobadilla, \emph{{Fast evaluation of
  Feynman integrals for Monte Carlo generators}},
  \href{https://doi.org/10.1007/JHEP09(2025)210}{\emph{JHEP} {\bfseries 09}
  (2025) 210} [\href{https://arxiv.org/abs/2507.12548}{{\ttfamily
  2507.12548}}].

\bibitem{Pozzoli:2026eiu}
M.~Pozzoli and W.~J. Torres~Bobadilla, \emph{{First look at the evaluation of
  two-loop Feynman integrals for radiative return processes}},
  \href{https://arxiv.org/abs/2607.03343}{{\ttfamily 2607.03343}}.

\bibitem{Kollatzsch:2026TI}
{Sophie Kollatzsch on behalf of the McMule team}, \emph{{McMule update}},  in
  \emph{{9th Plenary Workshop of the Muon $g-2$ Theory Initiative}}, (Storrs),
  2026,
  \href{https://indico.phys.uconn.edu/event/3/contributions/114/}{https://indico.phys.uconn.edu/event/3/contributions/114/}.

\bibitem{Fang:2026dbs}
Y.~Fang, S.~G{\"u}ndogdu, S.~Kollatzsch, M.~Rocco, A.~Signer and Y.~Ulrich,
  \emph{{NNLO initial-state corrections to $e^+\,e^-\to
  \mu^+\,\mu^-\,\gamma$}},  \href{https://arxiv.org/abs/2610.09680}{{\ttfamily
  2610.09680}}.

\bibitem{Colangelo:2015ama}
G.~Colangelo, M.~Hoferichter, M.~Procura and P.~Stoffer, \emph{{Dispersion
  relation for hadronic light-by-light scattering: theoretical foundations}},
  \href{https://doi.org/10.1007/JHEP09(2015)074}{\emph{JHEP} {\bfseries 09}
  (2015) 074} [\href{https://arxiv.org/abs/1506.01386}{{\ttfamily
  1506.01386}}].

\bibitem{Kaziukenas:2025gggpp}
E.~Kaziuk{\.e}nas, \emph{{Dispersive definition of
  $\gamma^*\gamma^*\gamma\to\pi^+\pi^-$ for muon $g-2$ applications}},  in
  \emph{{RadioMonteCarLow 2 Satellite 2025}}, (Liverpool), 2025,
  \href{https://indico.ph.liv.ac.uk/event/2169/contributions/10215/}{https://indico.ph.liv.ac.uk/event/2169/contributions/10215/}.

\bibitem{Sakurai:1972wk}
J.~J. Sakurai and D.~Schildknecht, \emph{{Generalized vector dominance and
  inelastic electron - proton scattering}},
  \href{https://doi.org/10.1016/0370-2693(72)90300-0}{\emph{Phys. Lett. B}
  {\bfseries 40} (1972) 121}.

\bibitem{Ignatov:2022iou}
F.~Ignatov and R.~N. Lee, \emph{{Charge asymmetry in $e^+e^-\to\pi^+\pi^-$
  process}}, \href{https://doi.org/10.1016/j.physletb.2022.137283}{\emph{Phys.
  Lett. B} {\bfseries 833} (2022) 137283}
  [\href{https://arxiv.org/abs/2204.12235}{{\ttfamily 2204.12235}}].

\bibitem{PetitRosas:2026iuq}
P.~Petit~Ros{\`a}s, O.~Shekhovtsova and W.~J. Torres~Bobadilla,
  \emph{{Radiative return meets GVMD}},
  \href{https://doi.org/10.1016/j.physletb.2026.140864}{\emph{Phys. Lett. B}
  {\bfseries 881} (2026) 140864}
  [\href{https://arxiv.org/abs/2603.13171}{{\ttfamily 2603.13171}}].

\bibitem{CarloniCalame:2026vfc}
C.~M. Carloni~Calame, M.~Ghilardi, A.~Gurgone, G.~Montagna, M.~Moretti,
  O.~Nicrosini et~al., \emph{{Structure-dependent radiative corrections to
  \(\left. e^{+}e^{-}\rightarrow\pi^{+}\pi^{-}\gamma \right.\) in the GVMD
  approach}}, \href{https://doi.org/10.1016/j.physletb.2026.140666}{\emph{Phys.
  Lett. B} {\bfseries 879} (2026) 140666}
  [\href{https://arxiv.org/abs/2603.28621}{{\ttfamily 2603.28621}}].

\bibitem{Delorme:2026qbm}
S.~Delorme, C.~Earnshaw, F.~Krauss, A.~Price, N.~Salone and A.~Si{\'o}dmok,
  \emph{{QED corrections to $e^+e^-\to\pi^+\pi^-(\gamma)$ with amplitude-level
  resummation}},  \href{https://arxiv.org/abs/2611.07067}{{\ttfamily
  2611.07067}}.

\bibitem{Colangelo:2014dfa}
G.~Colangelo, M.~Hoferichter, M.~Procura and P.~Stoffer, \emph{{Dispersive
  approach to hadronic light-by-light scattering}},
  \href{https://doi.org/10.1007/JHEP09(2014)091}{\emph{JHEP} {\bfseries 09}
  (2014) 091} [\href{https://arxiv.org/abs/1402.7081}{{\ttfamily 1402.7081}}].

\bibitem{Colangelo:2022lzg}
G.~Colangelo, M.~Hoferichter, J.~Monnard and J.~R. de~Elvira, \emph{{Radiative
  corrections to the forward-backward asymmetry in $e^+e^-\to\pi^+\pi^-$}},
  \href{https://doi.org/10.1007/JHEP08(2022)295}{\emph{JHEP} {\bfseries 08}
  (2022) 295} [\href{https://arxiv.org/abs/2207.03495}{{\ttfamily
  2207.03495}}].

\bibitem{Fang:2025mhn}
Y.~Fang, S.~Kollatzsch, M.~Rocco, A.~Signer, Y.~Ulrich and M.~Zoller,
  \emph{{Disperon QED}},
  \href{https://doi.org/10.21468/SciPostPhys.20.4.116}{\emph{SciPost Phys.}
  {\bfseries 20} (2026) 116}
  [\href{https://arxiv.org/abs/2512.10709}{{\ttfamily 2512.10709}}].

\bibitem{CarloniCalame:2026gdm}
C.~M. Carloni~Calame, M.~Ghilardi, A.~Gurgone, G.~Montagna, M.~Moretti,
  O.~Nicrosini et~al., \emph{{Next-to-leading order corrections to radiative
  pion pair production via dispersion relations}},
  \href{https://doi.org/10.1103/8zd9-sytq}{\emph{Phys. Rev. D} {\bfseries 114}
  (2026) 076022} [\href{https://arxiv.org/abs/2607.24642}{{\ttfamily
  2607.24642}}].

\bibitem{Czyz:2003ue}
H.~Czyz, A.~Grzelinska, J.~H. Kuhn and G.~Rodrigo, \emph{{The Radiative return
  at Phi and B factories: FSR at next-to-leading order}},
  \href{https://doi.org/10.1140/epjc/s2004-01605-0}{\emph{Eur. Phys. J. C}
  {\bfseries 33} (2004) 333}
  [\href{https://arxiv.org/abs/hep-ph/0308312}{{\ttfamily hep-ph/0308312}}].

\bibitem{Tracz:2018}
S.~Tracz, \emph{{Radiative corrections to hadrons-photons interactions}}, Ph.D.
  thesis, University of Silesia, 2018.

\bibitem{Campanario:2019mjh}
F.~Campanario, H.~Czy{\.z}, J.~Gluza, T.~Jeli{\'n}ski, G.~Rodrigo, S.~Tracz
  et~al., \emph{{Standard model radiative corrections in the pion form factor
  measurements do not explain the $a_\mu$ anomaly}},
  \href{https://doi.org/10.1103/PhysRevD.100.076004}{\emph{Phys. Rev. D}
  {\bfseries 100} (2019) 076004}
  [\href{https://arxiv.org/abs/1903.10197}{{\ttfamily 1903.10197}}].

\bibitem{Punzi:2026TI}
{Lorenzo Punzi on behalf of the KLOE-2 Collaboration}, \emph{{Status of the
  KLOE-NXT Analysis}},  in \emph{{9th Plenary workshop of the Muon $g-2$ Theory
  Initiative 2026}}, (University of Connecticut), 2026,
  \href{https://indico.phys.uconn.edu/event/3/contributions/129}{https://indico.phys.uconn.edu/event/3/contributions/129}.

\bibitem{KLOE:private}
F.~Ignatov and G.~Venanzoni, \emph{{Private communications}},  2026.

\bibitem{Hoferichter:2013ama}
M.~Hoferichter, G.~Colangelo, M.~Procura and P.~Stoffer, \emph{{Virtual
  photon-photon scattering}},
  \href{https://doi.org/10.1142/S2010194514604001}{\emph{Int. J. Mod. Phys.
  Conf. Ser.} {\bfseries 35} (2014) 1460400}
  [\href{https://arxiv.org/abs/1309.6877}{{\ttfamily 1309.6877}}].

\bibitem{Hoferichter:2019nlq}
M.~Hoferichter and P.~Stoffer, \emph{{Dispersion relations for
  $\gamma^*\gamma^*\to\pi\pi$: helicity amplitudes, subtractions, and anomalous
  thresholds}}, \href{https://doi.org/10.1007/JHEP07(2019)073}{\emph{JHEP}
  {\bfseries 07} (2019) 073}
  [\href{https://arxiv.org/abs/1905.13198}{{\ttfamily 1905.13198}}].

\bibitem{Ludtke:2023hvz}
J.~L{\"u}dtke, M.~Procura and P.~Stoffer, \emph{{Dispersion relations for
  hadronic light-by-light scattering in triangle kinematics}},
  \href{https://doi.org/10.1007/JHEP04(2023)125}{\emph{JHEP} {\bfseries 04}
  (2023) 125} [\href{https://arxiv.org/abs/2302.12264}{{\ttfamily
  2302.12264}}].

\bibitem{Lymperiadou:2026gggpp}
E.~Lymperiadou, \emph{{Updates on dispersive analysis for
  $\gamma^*\gamma^*\gamma \to \pi\pi$}},  in \emph{{RadioMonteCarLow 2 Workshop
  2026}}, (Torino), 2026,
  \href{https://indico.global/event/16381}{https://indico.global/event/16381}.

\bibitem{Monnard:2021pvm}
J.~Monnard, \emph{{Radiative corrections for the two-pion contribution to the
  hadronic vacuum polarization contribution to the muon $g-2$}}, Ph.D. thesis,
  Bern U., 2021.

\bibitem{Hoferichter:2025rescatteringpp}
M.~Hoferichter, \emph{{Rescattering corrections to the pion Compton tensor }},
  in \emph{{RadioMonteCarLow 2 Satellite 2025}}, (Liverpool), 2025,
  \href{https://indico.ph.liv.ac.uk/event/2169/contributions/10214/}{https://indico.ph.liv.ac.uk/event/2169/contributions/10214/}.

\bibitem{Budassi:2024whw}
E.~Budassi, C.~M. Carloni~Calame, M.~Ghilardi, A.~Gurgone, G.~Montagna,
  M.~Moretti et~al., \emph{{Pion pair production in $e^+e^-$ annihilation at
  next-to-leading order matched to Parton Shower}},
  \href{https://doi.org/10.1007/JHEP05(2025)196}{\emph{JHEP} {\bfseries 05}
  (2025) 196} [\href{https://arxiv.org/abs/2409.03469}{{\ttfamily
  2409.03469}}].

\bibitem{Banerjee:2020rww}
P.~Banerjee, T.~Engel, A.~Signer and Y.~Ulrich, \emph{{QED at NNLO with
  McMule}}, \href{https://doi.org/10.21468/SciPostPhys.9.2.027}{\emph{SciPost
  Phys.} {\bfseries 9} (2020) 027}
  [\href{https://arxiv.org/abs/2007.01654}{{\ttfamily 2007.01654}}].

\bibitem{McMule:manual}
\mcmule{} Team, ``\mcmule{} manual.''
  \url{https://doi.org/10.5281/zenodo.6046769}.

\bibitem{Colangelo:2017fiz}
G.~Colangelo, M.~Hoferichter, M.~Procura and P.~Stoffer, \emph{{Dispersion
  relation for hadronic light-by-light scattering: two-pion contributions}},
  \href{https://doi.org/10.1007/JHEP04(2017)161}{\emph{JHEP} {\bfseries 04}
  (2017) 161} [\href{https://arxiv.org/abs/1702.07347}{{\ttfamily
  1702.07347}}].

\bibitem{Aoyama:2020ynm}
T.~Aoyama et~al., \emph{{The anomalous magnetic moment of the muon in the
  Standard Model}},
  \href{https://doi.org/10.1016/j.physrep.2020.07.006}{\emph{Phys. Rept.}
  {\bfseries 887} (2020) 1} [\href{https://arxiv.org/abs/2006.04822}{{\ttfamily
  2006.04822}}].

\bibitem{Yennie:1961ad}
D.~R. Yennie, S.~C. Frautschi and H.~Suura, \emph{{The infrared divergence
  phenomena and high-energy processes}},
  \href{https://doi.org/10.1016/0003-4916(61)90151-8}{\emph{Annals Phys.}
  {\bfseries 13} (1961) 379}.

\bibitem{Engel:2019nfw}
T.~Engel, A.~Signer and Y.~Ulrich, \emph{{A subtraction scheme for massive
  QED}}, \href{https://doi.org/10.1007/JHEP01(2020)085}{\emph{JHEP} {\bfseries
  01} (2020) 085} [\href{https://arxiv.org/abs/1909.10244}{{\ttfamily
  1909.10244}}].

\bibitem{Buccioni:2017yxi}
F.~Buccioni, S.~Pozzorini and M.~Zoller, \emph{{On-the-fly reduction of open
  loops}}, \href{https://doi.org/10.1140/epjc/s10052-018-5562-1}{\emph{Eur.
  Phys. J. C} {\bfseries 78} (2018) 70}
  [\href{https://arxiv.org/abs/1710.11452}{{\ttfamily 1710.11452}}].

\bibitem{Buccioni:2019sur}
F.~Buccioni, J.-N. Lang, J.~M. Lindert, P.~Maierh{\"{o}}fer, S.~Pozzorini,
  H.~Zhang et~al., \emph{{OpenLoops 2}},
  \href{https://doi.org/10.1140/epjc/s10052-019-7306-2}{\emph{Eur. Phys. J. C}
  {\bfseries 79} (2019) 866}
  [\href{https://arxiv.org/abs/1907.13071}{{\ttfamily 1907.13071}}].

\bibitem{Kollatzsch:2026ubi}
S.~Kollatzsch, \emph{{Beyond QED: Electroweak and hadronic extensions of
  McMule}},  in \emph{{17th International Symposium on Radiative Corrections:
  Applications of Quantum Field Theory to Phenomenolog}}, 3, 2026,
  \href{https://arxiv.org/abs/2603.09443}{{\ttfamily 2603.09443}}.

\bibitem{Colangelo:2018mtw}
G.~Colangelo, M.~Hoferichter and P.~Stoffer, \emph{{Two-pion contribution to
  hadronic vacuum polarization}},
  \href{https://doi.org/10.1007/JHEP02(2019)006}{\emph{JHEP} {\bfseries 02}
  (2019) 006} [\href{https://arxiv.org/abs/1810.00007}{{\ttfamily
  1810.00007}}].

\bibitem{McMule:data}
\mcmule{} Team, ``\mcmule{} dataset.''
  \url{https://doi.org/10.5281/zenodo.8188752}.

\bibitem{Hoferichter:2026rescattering}
M.~Hoferichter, \emph{{Thoughts on rescattering corrections}},  in
  \emph{{RadioMonteCarLow 2 Workshop 2026}}, (Torino), 2026,
  \href{https://indico.global/event/16381}{https://indico.global/event/16381}.

\bibitem{Hoferichter:note2026}
M.~Hoferichter, \emph{{Towards mixed initial- and final-state corrections to
  $e^+e^- \to \pi^+ \pi^- \gamma$ beyond scalar QED}},  2026.

\bibitem{Denner:2002ii}
A.~Denner and S.~Dittmaier, \emph{{Reduction of one loop tensor five point
  integrals}}, \href{https://doi.org/10.1016/S0550-3213(03)00184-6}{\emph{Nucl.
  Phys. B} {\bfseries 658} (2003) 175}
  [\href{https://arxiv.org/abs/hep-ph/0212259}{{\ttfamily hep-ph/0212259}}].

\bibitem{Bern:1992em}
Z.~Bern, L.~J. Dixon and D.~A. Kosower, \emph{{Dimensionally regulated one loop
  integrals}}, \href{https://doi.org/10.1016/0370-2693(93)90400-C}{\emph{Phys.
  Lett. B} {\bfseries 302} (1993) 299}
  [\href{https://arxiv.org/abs/hep-ph/9212308}{{\ttfamily hep-ph/9212308}}].

\bibitem{Bern:1993kr}
Z.~Bern, L.~J. Dixon and D.~A. Kosower, \emph{{Dimensionally regulated pentagon
  integrals}}, \href{https://doi.org/10.1016/0550-3213(94)90398-0}{\emph{Nucl.
  Phys. B} {\bfseries 412} (1994) 751}
  [\href{https://arxiv.org/abs/hep-ph/9306240}{{\ttfamily hep-ph/9306240}}].

\bibitem{Ellis:2007qk}
R.~K. Ellis and G.~Zanderighi, \emph{{Scalar one-loop integrals for QCD}},
  \href{https://doi.org/10.1088/1126-6708/2008/02/002}{\emph{JHEP} {\bfseries
  02} (2008) 002} [\href{https://arxiv.org/abs/0712.1851}{{\ttfamily
  0712.1851}}].

\end{thebibliography}\endgroup

\end{document}